\documentclass[10pt]{article}

\usepackage{amsmath}

\usepackage{array}

\usepackage{appendix}

\usepackage{tocloft}                   

\usepackage{graphicx}

\usepackage{amsfonts}

\usepackage{amssymb}

\usepackage{mathrsfs}

\usepackage{yfonts}

\usepackage{euscript}

\usepackage{centernot}                 

\usepackage{ifsym}                     

\usepackage{lmodern}                   

\usepackage{upgreek}

\usepackage{mathtools}

\usepackage{color}

\usepackage{slantsc}
\usepackage{calligra}

\usepackage{bbold}          

\usepackage[T1]{fontenc}

\usepackage{epsf}

\usepackage{latexsym}

\usepackage{tipa}

\usepackage{makeidx}

\makeindex

\def\b{\begin{equation}}

\def\e{\begin{equation}}

\def\be{\begin{equation}}              

\def\ee{\end{equation}}

\def\beq{\begin{equation}}

\def\eeq{\end{equation}}

\def\bea{\begin{eqnarray}}

\def\eea{\end{eqnarray}}

\def\half{\mbox{$\frac{1}{2}$}}

\def\m{\mbox{ }}

\def\mma {\m , \m \m }

\def\!{\hspace{-1.6667em}}

\def\n{\noindent}

\def\u{\underline}

\def\slLambda{\mathit{\Lambda}}                   

\def\uix{\u{x}}

\def\biP{\mbox{\boldmath$P$}}

\def\biQ{\mbox{\boldmath$Q$}}

\def\sbiN{\mbox{\scriptsize\boldmath$N$}}

\def\sbiQ{\mbox{\scriptsize\boldmath$Q$}}

\def\boldeta{\mbox{\boldmath$\eta$}}                

\def\mA{\mbox{A}}  

\def\mB{\mbox{B}}  

\def\mE{\mbox{E}}                        

\def\mF{\mbox{F}}

\def\mI{\mbox{I}}                        

\def\mL{\mbox{L}}

\def\mM{\mbox{M}}                        

\def\mN{\mbox{N}}

\def\mP{\mbox{P}}

\def\mS{\mbox{S}}                        

\def\mT{\mbox{T}} 

\def\mV{\mbox{V}}

\def\mb{\mbox{b}}

\def\mh{\mbox{h}}

\def\mo{\mbox{o}}

\def\mp{\mbox{p}}

\def\ms{\mbox{s}}

\def\mt{\mbox{t}}

\def\bh{\u{\u{\mbox{h}}}  }            

\def\sbh{\u{\u{\mbox{\scriptsize h}}}  }     

\def\bC{\mbox{\bf C}}                    

\def\bM{\mbox{\bf M}}

\def\bQ{\mbox{\bf Q}}

\def\bd{\mbox{\bf d}}

\def\bh{\mbox{\bf h}}

\def\bp{\mbox{\bf p}}

\def\bdelta{\mbox{\boldmath$\delta$}}

\def\bupSigma{\mbox{\boldmath$\Sigma$}}                 
\def\sbupSigma{\mbox{\scriptsize\boldmath$\Sigma$}}     
\def\lbupSigma{\mbox{\large \boldmath$\Sigma$}}          

\def\sbSigma{\mbox{\scriptsize\boldmath$\Sigma$}}

\def\bcalK{\mbox{\boldmath ${\cal K}$}} 

\def\bcalD{\mbox{\boldmath ${\cal D}$}} 

\def\bcalR{\mbox{\boldmath ${\cal R}$}}

\def\bicM{\mbox{\boldmath${\cal M}$}}

\def\biscM{\mbox{\scriptsize\boldmath${\cal M}$}}

\def\fc{\mbox{\sffamily c}}

\def\fg{\mbox{\tt g}}                          

\def\fm{\mbox{\sffamily m}}

\def\fs{\mbox{\sffamily s}}

\def\ft{\mbox{\sffamily t}}                       

\def\fz{\mbox{\sffamily z}}

\def\fA{\mbox{\sffamily A}}           

\def\bfg{\mbox{\bf\sffamily g}}

\def\fF{\mbox{\sffamily F}}

\def\fG{\mbox{\sffamily G}}

\def\fM{\mbox{\sffamily M}}

\def\fN{\mbox{\sffamily N}}

\def\fP{\mbox{\sffamily P}}

\def\fQ{\mbox{\sffamily Q}}

\def\fT{\mbox{\sffamily T}}

\def\fW{\mbox{\sffamily W}}

\def\fZ{\mbox{\sffamily Z}}

\def\cA{{\mathscr A}}

\def\cF{{\mathscr F}}

\def\cH{{\mathscr H}}

\def\cJ{{\mathscr J}}

\def\cL{{\mathscr L}}

\def\cR{{\mathscr R}}

\def\cV{{\mathscr V}}

\def\cW{{\mathscr W}}

\def\sa{\mbox{\scriptsize a}}

\def\scc{\mbox{\scriptsize c}}

\def\sd{\mbox{\scriptsize d}}

\def\se{\mbox{\scriptsize e}}

\def\sg{\mbox{\scriptsize g}} 

\def\sh{\mbox{\scriptsize h}} 

\def\si{\mbox{\scriptsize i}}

\def\sll{\mbox{\scriptsize l}}  

\def\sm{\mbox{\scriptsize m}}

\def\sn{\mbox{\scriptsize n}} 

\def\so{\mbox{\scriptsize o}}

\def\sq{\mbox{\scriptsize q}}

\def\sr{\mbox{\scriptsize r}}

\def\sss{\mbox{\scriptsize s}}  

\def\st{\mbox{\scriptsize t}}

\def\su{\mbox{\scriptsize u}}

\def\sv{\mbox{\scriptsize v}}

\def\sx{\mbox{\scriptsize x}}

\def\sA{\mbox{\scriptsize A}} 

\def\sB{\mbox{\scriptsize B}}

\def\sF{\mbox{\scriptsize F}}

\def\sG{\mbox{\scriptsize G}}

\def\sI{\mbox{\scriptsize I}}

\def\sM{\mbox{\scriptsize M}} 

\def\sN{\mbox{\scriptsize N}} 

\def\sO{\mbox{\scriptsize O}}

\def\sR{\mbox{\scriptsize R}}

\def\sS{\mbox{\scriptsize S}}

\def\sW{\mbox{\scriptsize W}}

\def\sfc{\mbox{\sffamily{\scriptsize c}}}     

\def\sfg{\mbox{\sffamily{\scriptsize g}}}     

\def\sfz{\mbox{\sffamily{\scriptsize z}}}     

\def\sfA{\mbox{\sffamily{\scriptsize A}}}     

\def\sfB{\mbox{\sffamily{\scriptsize B}}}     

\def\sfC{\mbox{\sffamily{\scriptsize C}}}     

\def\sfY{\mbox{\sffamily{\scriptsize Y}}}      

\def\sfZ{\mbox{\sffamily{\scriptsize Z}}}      

\def\sbh{\mbox{{\bf \scriptsize h}}}

\def\sbd{\mbox{{\bf \scriptsize d}}}

\def\sbM{\mbox{{\bf \scriptsize M}}}

\def\sbN{\mbox{{\bf \scriptsize N}}}

\def\sbP{\mbox{{\bf \scriptsize P}}}

\def\sbfm{\mbox{\bf \scriptsize\sffamily m}}

\def\sbfg{\mbox{\bf \scriptsize\sffamily g}}

\def\bfg{\mbox{\bf \sffamily g}}

\def\sbfM{\mbox{\bf \scriptsize\sffamily M}}

\def\sbfP{\mbox{\bf \scriptsize\sffamily P}}

\def\sbfQ{\mbox{\bf \scriptsize\sffamily Q}}

\def\sbcL{\mbox{\boldmath \scriptsize ${\cal L}$}}

\def\sbcS{\mbox{\boldmath \scriptsize ${\cal S}$}}

\def\usF{\u{\mbox{\scriptsize F}}}

\def\ta{\mbox{\tiny a}}

\def\tl{\mbox{\tiny l}}

\def\to{\mbox{\tiny o}}

\def\ttt{\mbox{\tiny t}}   

\def\tT{\mbox{\tiny T}}

\def\bfg{\mbox{{\bf \sffamily g}}}                                    

\def\bfQ{\mbox{{\bf \sffamily Q}}}                                    

\def\bsfc{\mbox{{\bf\scriptsize \sffamily c}}}                        

\def\bfP{\mbox{{\bf \sffamily P}}}                                    

\def\bfM{\mbox{{\bf \sffamily M}}}                                    

\def\btcF{\mbox{{\boldmath \tiny${\cal F}$}}}                               

\def\bscM{\mbox{{\bf \scriptsize${\cal M}$}}}                               

\def\bscF{\mbox{{\boldmath \scriptsize${\cal F}$}}}                               

\def\sbfQ{\mbox{{\bf \scriptsize\sffamily Q}}}                               %

\def\bfm{\mbox{{\bf \sffamily m}}}                                    %

\def\bfc{\mbox{{\bf \sffamily c}}}                                    

\def\bfp{\mbox{{\bf \sffamily p}}}                                    

\def\cr{\mbox{\scriptsize{\bf $\m  \times \m $}}}

\def\sumi2{\sum\mbox{}_{\mbox{}_{\mbox{\scriptsize $i$=1}}}^2}

\def\sumi3{\sum\mbox{}_{\mbox{}_{\mbox{\scriptsize $i$=1}}}^3}

\def\sumABcycles3{\sum\mbox{}_{\mbox{}_{\mbox{\scriptsize cycles $A,B$=1}}}^{3}}

\def\sumCDcycles3{\sum\mbox{}_{\mbox{}_{\mbox{\scriptsize cycles $C,D$=1}}}^{3}}

\def\sumj3{\sum\mbox{}_{\mbox{}_{\mbox{\scriptsize $j$=1}}}^3}

\def\sumk3{\sum\mbox{}_{\mbox{}_{\mbox{\scriptsize $k$=1}}}^3}

\def\prodiA1{\prod\mbox{}_{\mbox{}_{\mbox{\scriptsize $i$=1}}}^{A - 1}}

\def\d{\textrm{d}}                                                  

\def\pa{\partial}                                                   

\def\ordial{\bd\hspace{-0.088in}\pa}                                

\def\partional{\bdelta\hspace{-0.08in}\pa}                          

\def\sordial{{\sbd\hspace{-0.075in}\mbox{\scriptsize$\pa$}}}        

\def\Circ{\mbox{\Large$\circ$}}                                     

\def\Last{\mbox{\Large$\ast$}}                                      

\def\es{\m = \m}

\def\:={\m := \m}

\def\=:{\m =: \m}

\def\Abs{\mbox{\Large $\mathfrak{a}$}\mb\ms}                         

\def\Int{\mbox{\Large $\mathfrak{i}$}}                         

\def\FrT{\mathfrak{T}}                                         

\def\sFrS{\mbox{\large$\mathfrak{s}$}}                         

\def\lFrm{\mbox{\large$\mathfrak{m}$}}                         
\def\lFrg{\mbox{\Large$\mathfrak{g}$}}                         
\def\nFrg{\mbox{\large$\mathfrak{g}$}}                         
\def\FrT{\mbox{\boldmath$\mathfrak{T}$}}                       
\def\Hilb{\mbox{{\boldmath$\mathfrak{H}$}ilb}}                 
\def\Ban{\mbox{{\boldmath$\mathfrak{B}$}an}}                   

\def\Fre{\mbox{{\boldmath$\mathfrak{f}$}re}}                   

\def\Frc{\mbox{\Large $\mathfrak{c}$}}                         
\def\FrL{\mbox{\boldmath$\mathfrak{L}$}}                       
\def\lft{\mbox{\Large \sffamily t}}                        

\def\lt{\mbox{\Large $t$}}                                 

\def\lmt{\mbox{\Large t}}                                  

\def\scC{\mbox{\scriptsize ${\cal C}$}}                    
\def\scG{\mbox{\scriptsize ${\cal G}$}}                    

\def\scH{\mbox{\scriptsize ${\cal H}$}}                    

\def\scM{\mbox{\scriptsize ${\cal M}$}}                    

\def\bLin{\sbcL\mbox{\bf in}} 

\def\Chronos{\scC\mbox{hronos}}                            

\def\bShuffle{\sbcS\mbox{\bf huffle}} 

\def\FrQ{\mbox{\Large $\mathfrak{q}$}}                               
\def\Phase{\mbox{{\boldmath$\mathfrak{P}$}hase}}                     

\def\bFrR{\mbox{\boldmath$\mathfrak{R}$}}                            
\def\Rig-Phase{\bFrR\mbox{ig-}\Phase}                                
                                                                													   
\def\sFS{\mbox{\large$\mathfrak{s}$}}                                      
	
\def\bFrR{\mbox{\boldmath$\mathfrak{R}$}}                            

\def\bFrR{\mbox{\boldmath$\mathfrak{R}$}}                            

\def\1mat{\u{\u{1}}}                                                 

\def\Sym{\bFrS\mbox{ym}}                                             
\def\sSym{\sbFrS\mbox{\scriptsize ym}}                               
\def\Positive-Modespace{\mbox{{\boldmath$\mathfrak{M}$}odespace$^+$}}

\def\POSITIVE-MODESPACE{\mbox{{\boldmath$\mathfrak{M}$}ODESPACE$^+$}}
                                                                                                                             														
\def\bFrS{\mbox{\Large $\mathfrak{s}$}}                              

\def\sbFrS{\mbox{\large\boldmath$\mathfrak{s}$}}                     
\def\blFrS{\mbox{\LARGE $\mathfrak{s}$}}                             
\def\Riem{\bFrR\mbox{iem}}                                           
\def\Superspace{\bFrS\mbox{uperspace}}                               
\def\lSuperspace{\blFrS\mbox{uperspace}}                             

\def\RIEM{\bFrR\mbox{IEM}}                                           

\def\Sca{\bFrS\mbox{ca}}                                             

\def\sSuperspacetime{\sbFrS\mbox{\scriptsize uperspacetime}}         

\def\lE{\mbox{\Large E}}

\def\Kin-Hilb{\mbox{{\boldmath$\mathfrak{K}$}in-\Hilb}}                     

\def\Mid-Hilb{\mbox{{\boldmath$\mathfrak{M}$}id-\Hilb}}                     

\def\Dyn-Hilb{\mbox{{\boldmath$\mathfrak{D}$}yn-\Hilb}}                     

\def\5Star{\mbox{\Large$\star$}}              

\begin{document}

\begin{center}

\Huge{\bf A LOCAL RESOLUTION OF}

\vspace{.1in}

\normalsize

\Huge{\bf THE PROBLEM OF TIME}

\vspace{.15in}

\Large{\bf VI. Combining Temporal and Configurational Relationalism} 

\m

\Large{\bf for Field Theories and GR}

\vspace{.1in}

{\large \bf E.  Anderson} 

\vspace{.15in}

{\large \sl based on calculations done at Peterhouse, Cambridge} 

\end{center}

\begin{abstract}

We next combine Temporal and Configurational Relationalism's resolution for Field Theory, including in particular for GR. 
The current Article also provides the finite-and-field theory portmanteau notation, 
by which the rest of this series' reworking of the Principles of Dynamics can be presented concurrently for Finite Theories and Field Theories.
GR's Riem, superspace, and thin sandwich are also further outlined in support of the rest of this Series.  

\end{abstract}

\section{Introduction}

In this sixth Article on the Problem of Time \cite{Battelle, DeWitt67, Dirac, K81, K91, K92, I93, K99, APoT, FileR, APoT2, AObs, APoT3, ALett, 
ABook, A-CBI, I, II, III, IV}, we combine \cite{FEPI, FileR, ABook, MBook} Article I's Temporal Relationalism and Article II's Configurational Relationalism 
for Field Theories (a more advanced counterpart to Article V's Finite Theory arena).

\m 

\n The current Article also provides the finite-and-field theory portmanteau notation in Sec 2 
by which the rest of this series' reworking of the Principles of Dynamics can be presented concurrently for Finite Theories and Field Theories.
Sec 3 next outlines unreduced field theoretic configuration spaces. 
This is followed by Sec 4's Principles of Dynamics (PoD) for Field Theory, 
provided for comparison with Sec 5's rendition of Temporal Relationalism implementing Principles of Dynamics (TRiPoD) for Field Theory.
We then provide Electromagnetism, pure GR, GR with minimally-coupled scalar field, Electromagnetism and Yang--Mills Theory, and Strong Gravity, 
as examples in Secs 5 to 9 respectively. 
Sec 6 includes further outline of GR's reduced configuration space, Superspace \cite{Battelle, DeWitt67, DeWitt70, Fischer70, FM96, Giu09, Giu15}.

\section{Field and portmanteau variables}\label{FVs}

Joint treatment of finite and field-theoretic models begins in this Section, by declaring portmanteau notation.
$\bfQ$ is the portmanteau configuration of finite theories' configuration $\biQ$ and Field Theories' $\bQ = \bQ(\uix)$ configuration.  
Each of these carries the $\fA$ multi-index, over one or both of particle or continuous extended object species.

\m 

\n{\bf Remark 1} Field theories have $\pa$ in place of $\d$, and functional derivatives $\delta$ in place of $\pa$ in their PoD.

\m  

\n{\bf Definition 1} Let us also introduce two `{\it derivative portmanteaux}': 
{\it ordial derivatives}    $\ordial$:    ordinary--partial derivative portmanteau, and 
{\it partional derivatives}	$\partional$: partial--functional derivative portmanteau.     

\m 

\n{\bf Remark 2} In passing, let us rename the Functional Evolution Problem as the Partional Evolution Problem to jointly cover Finite and Field Theories. 
   
\m  

\n{\bf Remark 3} We take the field theoretic kinetic metric to be {\it ultralocal} -- i.e.\ having no derivative dependence --
a mathematical simplicity which happens to hold over the entirety of the standardly accepted fundamental theories of Physics.

\m 

\n{\bf Remark 4} We use the kinetic metric portmanteau notation  
\be 
{\fM}_{\sfA\sfB}  =  M_{\sfA\sfB}(\biQ) \m \mbox{ (finite)} \m \mbox{ or } \m   {\mM}_{\sfA\sfB}({\bQ}(\uix)) \m \mbox{ (field) } \m . 
\ee 
We denote the corresponding determinant, inverse, inner product and `norm' (indefiniteness allowed) by 
$\fM$, $\fN^{\sfA\sfB}$, $(\m , \m)_{\sbfM}\mbox{}$ and $||\m||_{\sbfM}\mbox{}$ respectively.
$\Circ$ is now considered to take the {\it ordial derivative} form 
\be 
\frac{\ordial}{\ordial\lambda} \m :
\ee 
the portmanteau of ordinary and partial derivatives.

\m 

\n{\bf Structure 1} Our first aim is to build relational actions, in particular in a manner rich enough to include the case of full GR.\footnote{In fact, 
the first of these can be rewritten as $\mL(t; \sbiQ]$, which is a univariate functional due to $\d/\d t$ acting on the $\sbiQ$ to form the velocities.
However, this does not affect the types of derivatives that the theory has acting upon $\mL$, so it does not disrupt the portmanteau.
It makes sense to be able to talk about portmanteau derivatives because of Banach      Calculus \cite{AMP} 
                                                                        or Fr\'{e}chet Calculus \cite{Hamilton82} to standard calculus parallels.}
\be 
\cL\lfloor \dot{\bfQ}, \bfQ \rfloor = L(t, \biQ, \dot{\biQ}) \m \mbox{ (finite) } \m \mbox{ or } 
                                      {\cal L}(\uix, t, \dot{\bQ}; \bQ] \m \mbox{ (field: Lagrangian density)}  \m .
\ee 
The latter is taken to be ultralocal in the velocities for Field Theories.
One then obtains the relational action by integrating over the time portmanteau $\ft$ and the notion of space portmanteau.  

\m 

\n{\bf Remark 5} Compliance with Manifest Reparametrization Irrelevance does make the Lagrangian portmanteau in question {\sl look} somewhat unusual. 
I.e.\ it is not 
\be 
\mbox{of {\it difference-type form} } \m  \cL = \fT - \cV  \m ,
\ee 
but rather 
\be 
\mbox{of {\it product form} } \m \cL = 2 \sqrt{\fT \, \cW} \m .  
\ee 
For now, we take on trust that the potential factor portmanteau $\cW = E - \cV$, for potential energy $\cV$ and $\fT$ is the kinetic energy.  

\m 

\n{\bf Structure 2} To pass to a geometrical action presentation, we require rather 

\m 

\n 1) the kinetic arc element portmanteau 
$$ 
\ordial\fs \lfloor \bfQ, \ordial{\bfQ} \rfloor \mbox{  -- of the kinetic arc element    }    
\d s(\biQ, \d \biQ) \mbox{ for finite theories and the kinetic arc element density } 
$$
\be 
\pa \ms(\uix, \pa\bQ; \bQ] \mbox{ for Field Theories} \m .
\ee 
\n 2) The physical Jacobi arc element portmanteau 
$$ 
\ordial\cJ\lfloor {\bfQ}, \ordial\bfQ \rfloor \mbox{ -- of the Jacobi arc element }        
\d     J(\biQ, \d \biQ) \mbox{ for finite theories and the Jacobi arc element density }  
$$
\be
\pa{\cal J}(\uix, \pa {\bQ} (\uix); \bQ(\uix)] \mbox{ for Field Theories } .  
\ee 

\m 

\n{\bf Remark 6} As regards the nature of the geometries, these are now in fact infinite-dimensional generalizations of the previous Sections' Riemannian Geometry. 
Moreover, the local square root does not coincide with the DeWitt-type geometry, adding extra degeneracy and functional-based issues.
For convenience, let us still refer to such by the usual finite-dimensional geometries' nomenclature. 
I.e.\ we elevate names like `Riemannian' to be finite and field-theoretic portmanteaux of the usual finite version of the notion. 

\m 

\n{\bf Remark 7} In the case of Field Theories, Article II's definitions of inconsistent, trivial and relationally trivial 
are recast in terms of degrees of freedom per space point.
Care has to be taken now as regards nontrivial global degrees of freedom surviving.

\section{Unreduced configuration space geometry for Field Theory and GR}\label{GR-Config}

\subsection{Field Theory}\label{Q-Fields}

\n{\bf Structure 1} Scalar Field Theory's configuration space $\Sca$ the a space of scalar field values $\phi(\underline{x})$ 

\m 

\n{\bf Structure 2} Electromagnetism's configuration space is a space $\mbox{\boldmath$\Lambda^1$ of 1-forms $\mA_i(\underline{x})$}$.
             
\m 			 
			 
\n{\bf Structure 3} Yang--Mills Theory's configuration space is a larger space $\mbox{\boldmath$\Lambda$}$ of 1-forms $\mA_i^P(\underline{x})$.

\m 

\n{\bf Remark 1} $\Sca$ and $\mbox{\boldmath$\Lambda$}^1$ have implicit dependence $(\mathbb{R}^3)$ in many of their more standard uses.

\m 

\n{\bf Remark 2} In modelling the above in more detail, the square-integrable functions $\FrL^2$ provide one starting point.  
One can furthermore pass to e.g.\ the Fr\'{e}chet spaces of the next Section, which are subsequently useful in curved-space and GR-coupled versions. 
Here one has $\bupSigma$ dependence in place of $\mathbb{R}^3$.

\subsection{From Hilbert to Banach and Fr\'{e}chet spaces}\label{Hilb-Ban-Fre}

Consider the following ladder of increasingly general topological vector spaces which are infinite-$d$ function spaces \cite{AMP, TVS}.
        A {\it Hilbert space} $\Hilb$          is a complete inner product space, 
        a {\it Banach space}  $\Ban$           is a complete normed space,
and     a {\it Fr\'{e}chet space} $\Fre$ is a complete metrizable locally convex topological vector space \cite{Hamilton82}. 

\m 

\n{\bf Remark 1} Hilbert Spaces are the most familiar in Theoretical Physics due to their use in linear-PDE Fourier Analysis and in Quantum Theory.  

\m 

\n{\bf Remark 2} See \cite{AMP} for Calculus on Banach spaces. 
Functional Analysis has furthermore been extensively developed for Banach spaces \cite{AMP} 
Major results here include the {\it Hahn--Banach Theorem}, 
                           the {\it Uniform Boundedness Principle}, 
                       and the {\it Open Mapping Theorem}; see \cite{DS88} for details and proofs.

\m 

\n{\bf Remark 3} Treatment of GR configuration spaces moreover involves the even more general Fr\'{e}chet spaces. 
On the one hand, many substantial results in Functional Analysis -- in particular 1) to 3) -- 
furthermore carry over from Banach spaces to Fr\'{e}chet spaces \cite{Hamilton82}.
On the other hand, we caution that there is no longer in general an Inverse Function Theorem in this setting, 
though the {\it Nash--Moser Theorem} \cite{Hamilton82} is a replacement for this for a subclass of Fr\'{e}chet spaces.  
%
%
See \cite{Hamilton82} as regards Calculus on Fr\'{e}chet spaces more generally.  

\m 

\n Out finite--field portmanteau rests on the many parallels between standard Calculus and Banach or Fr\'{e}chet Calculus.

\subsection{Hilbert, Banach and Fr\'{e}chet Manifolds}\label{Inf-Manifolds}

\n{\bf Structure 1} Topological manifolds' local Euclideanness and ensuing $\mathbb{R}^p$-portion charts extend well to infinite-$d$ cases,   
for which the charts involve portions of Hilbert, Banach and Fr\'{e}chet spaces.
See e.g.\ \cite{Lang95, AMP, Hamilton82} for accounts of Hilbert, Banach and Fr\'{e}chet manifolds respectively. 
Banach manifolds are the limiting case as regards retaining a very wide range of analogies with finite manifolds.
      Fr\'{e}chet manifolds remain reasonably tractable \cite{AMP}, despite the loss in general of the Inverse Function Theorem, 
as do Fr\'{e}chet Lie groups \cite{Hamilton82}.

\m 

\n{\bf Structure 2} Finite manifolds' incorporation of differentiable structure also has an analogue in each of the above cases.
So e.g.\ one can consider differentiable functions and tangent vectors for each, 
and then apply multilinearity to set up versions for tensors of any other rank $(p, q)$ and symmetry type $S$. 
In particular, applying this construction to a Fr\'{e}chet manifold with tangent space 
\be 
\Fre(\Frc^{\infty})
\ee 
produces another Fr\'{e}chet space 
\be 
\Fre_{S(p, q)}(\Frc^{\infty}) \m . 
\ee

\subsection{Topology of $\Riem(\lbupSigma)$}\label{Riem-Top}

{\bf Structure 1} The space of Riemannian geometries $\Riem(\bupSigma)$ can be modelled as 
an open positive convex cone\footnote{`Cone' is here meant in a Linear Algebra sense: a space $\sFrS$  
\cite{Fischer70, RS}, that is not itself linear but obeys $\sFrS + \sFrS \subset \sFrS$ and $m \, \sFrS \subset \sFrS$ for $m \in \mathbb{R}_+$. 
See \cite{Fischer70} for more on this, alongside justification of the appropriateness of using Fr\'{e}chet spaces in this context.
It is not to be confused with this Series' more ubiquitous topological and geometrical uses of `cone'.}
in the Fr\'{e}chet space $\Fre_{\sSym(0, 2)}$($\Frc^{\infty}$) for $\sSym(0, 2)$ the symmetric rank-2 tensors.

\m

\n{\bf Structure 2} $\Riem(\bupSigma)$ can furthermore be equipped \cite{Fischer70} with a metric space notion of metric, Dist; 
this can additionally be chosen to be preserved under $Diff(\bupSigma)$. 
Thus $\Riem(\bupSigma$) is a metrizable topological space (XIV.1). 
Consequently $\Riem(\bupSigma$) obeys all the separation axioms -- including in particular Hausdorffness (XIV.1) -- and it is also paracompact (XIV.1). 
$\Riem(\bupSigma)$ is additionally second-countable \cite{Giu09} (XIV.1), 
and has an infinite-dimensional analogue of the locally Euclidean property as well; consequently a single type of chart suffices in this case. 
In this manner, $\Riem(\bupSigma)$ is a manifold that is infinite-dimensional in the sense of Fr\'{e}chet($\Frc^{\infty}$).

\subsection{Geometrical metric structure of $\Riem(\bupSigma)$}\label{Riem-Geom}

{\bf Structure 1} Infinite-$d$ manifolds can be equipped with connections and metrics \cite{AMP}. 
In the dynamical study of GR, $\Riem(\bupSigma)$ is usually taken to carry the infinite-dimensional indefinite Riemannian metric provided by GR's kinetic term, 
i.e.\ the inverse DeWitt supermetric $\bM$.  
More generally, one might consider other members of the family of {\it ultralocal supermetrics} \cite{Giu95b, KieferBook} 
\beq
\bM_{\beta} \m \mbox{ with components }  \m \mM_{\beta}^{abcd}  \:=  \sqrt{\mh}\{\mh^{ac}\mh^{bd} - w \, \mh^{ab}\mh^{cd}\}  \m .
\label{M-beta}
\eeq
These split into 3 cases: the positive-definite $w < 1/3$, the degenerate $w = 1/3$, and the indefinite (heuristically $\{-+++++\}^{\infty}$) $w > 1/3$.  
Ultralocality readily permits these to be studied pointwise; the more problematic degenerate case is usually dropped from such studies.
Pointwise, these supermetrics arise from positive-definite symmetric $3 \times 3$ matrices ($\mh_{ab}$ at that point. 
The 6-$d$ space of these is mathematically \cite{Giu95b} $\Sym^+(3,\mathbb{R})$, which is diffeomorphic to the homogeneous space 
\beq 
\frac{GL^+(3, \mathbb{R})}{ SO(3) } \m \, \cong \, \m \mathbb{R}_+ \times \mathbb{R}^5    \m . 
\label{Sym+} 
\eeq
This is the full Minisuperspace.  

\m 

\n{\bf Remark 1} Ultralocality also implies that this pointwise structure uplifts to $\Riem(\bupSigma)$.  
%
%
The scale-free part gives rise to 8 Killing vectors and the scale part to a homothety \cite{Giu95b}. 
The corresponding local Riemannian Geometry for this was studied by DeWitt \cite{DeWitt67}, including the form taken by the geodesics. 
This exhibits various global difficulties: curvature singularities and geodesic incompleteness \cite{Wald}.

\section{The standard Principles of Dynamics.  ii.  Field Theory}\label{PoD-Field}

\subsection{Space--time split GR version}\label{Field-PoD-4}

We next consider the ADM action \cite{ADM} (II.6, II.8); 
for inclusion of minimally-coupled scalars, Electromagnetism and Yang--Mills Theory, see Sec \ref{GR+Fund}.
GR's own momenta are (II.17), whereas the other field momenta are 
\beq
\uppi_{\sfZ}^{\uppsi} \:=  \frac{\updelta{\cal L}}{\updelta\dot{\uppsi}^{\sfZ}} \m . \m \m 
\eeq
Article II's versions of multiplier coordinates, cyclic coordinates, multiplier elimination, passage to the Routhian and to Hamiltonian carry over.  
ADM lapse $\upalpha$ and shift $\u{\upbeta}$ are now examples of multiplier coordinates. 
The bare GR Hamiltonian is zero, though of course there are constraints $\scH$ and $\u{\scM}$, giving Dirac's `extended' Hamiltonian.

\section{TRiPoD for Field Theory}\label{CR-Port}

\subsection{Jacobi--Mach formulation}\label{JM}

We continue to restrict our treatment to second-order physical systems, and now work in the absence of time at the primary level, as per Article I.  
Consequently, there is no derivative with respect to time and thus no notion of velocity $\dot{\bfQ}$ at the primary level.  
Instead, we use {\it change in configuration} 
\be 
\ordial \bfQ
\ee 
due to being open to resolving primary-level timelessness through Mach's Time Principle: with a secondary notion of time to be abstracted from change.
Thus in TRiPoD, {\it Machian variables} $(\bfQ, \ordial \bfQ)$ supplant the usual Principles of Dynamics's Lagrangian variables $(\bfQ, \dot{\bfQ})$. 

\m 

\n There is clearly no primary notion of kinetic energy; this has been supplanted by the {\it kinetic arc element} 
\be 
\ordial\fs  \es ||\d_{\sbfg} \bQ||_{\sbfM}  \m .  
\ee
All dynamical information is now contained within the {\it Jacobi arc element} 
\be 
\ordial\cJ(\bfQ, \ordial\bfQ)  \m ,
\ee 
which has supplanted the time-independent Lagrangian         
\be 
\cL(\bfQ, \dot{\bfQ})          \m .
\ee
The action  ${\cal S}$ is itself an {\sl unmodified} concept: it is already in TRi form, albeit now additionally bearing the relation 
\be 
{\cal S} = \int \ordial \cJ 
\ee 
to the TRiPoD formulation's Jacobi arc element $\ordial \cJ$.
Note the relation
\be 
\ordial \cJ  \es  \sqrt{2 \, \cW} \, \ordial \fs 
\ee 
for $\cW = \cW(\bfQ)$ the usual potential factor, so the kinetic and Jacobi arc elements are related by a conformal transformation.  
In terms of $\ordial \cJ$, Dynamics has been cast in the form of a {\it geodesic     principle} \cite{B94I}, or, 
in terms of $\ordial \fs$                                     as a {\it parageodesic principle} along the lines of Misner \cite{Magic}.

\m 

\n We next apply the Calculus of Variations to obtain the equations of motion such that ${\cal S}$ is stationary with respect to the $\bfQ$.
See Sec \ref{FENoS} for comments on the particular form taken by this variation.  
The resulting equations of motion the `{\it Jacobi--Mach equations}', 
\be
\ordial \left\{ \frac{\partional \, \ordial \cJ}{\partional \, \ordial \bfQ} \right\} - \frac{\partional \, \ordial \cJ}{\partional \bfQ}  \es  0   \m , 
\label{JME}
\ee
in place of the usual Principles of Dynamics's Euler--Lagrange equations.

\m

\n The Jacobi--Mach equations also admit three simplified cases.

\m

\n 1) {\it Lagrange multiplier coordinates} $\bfm \, \subseteq \, \bfQ$ are such that $\ordial \cJ$ is independent of $\ordial \bfm$, 
$$
\frac{\partional \, \ordial \cJ}{\partional \, \ordial \bfm}  \es  0  \m .
$$
The corresponding Jacobi--Mach equation is 
\be
\frac{\partional \, \ordial \cJ}{\partional \bfm}  \es  0             \m .
\label{lmel-2}
\ee
2) {\it Cyclic coordinates} $\bfc \, \subseteq \, \bfQ$  are such that $\ordial \cJ$ is independent of $\bfc$,  
\be 
\frac{\partional \, \ordial \cJ}{\partional \bfc}  \es 0              \m ,
\ee 
while still featuring $\ordial \bfc$: the corresponding {\it cyclic differential}.\footnote{To avoid confusion, 
`cyclic' in `cyclic differential' just means the same as `cyclic' in cyclic velocity. 
So nothing like `exact differential' or `cycle' in Algebraic Topology -- which in de Rham's case is tied to differentials -- is implied here.\label{Cyclic}} 
%
The corresponding Jacobi--Mach equation is 
\be
\frac{\partional \, \ordial \cJ}{\partional \, \ordial \bfc}  \es  \mbox{\bf const}  \m .
\label{cyclic-vel-2}
\ee
3) {\it The energy integral type simplification}.  
$\ordial \cJ$ is independent of what was previously regarded as `the independent variable $\ft$', whereby one Jacobi--Mach equation may be supplanted by the first integral
\be
\ordial \cJ -  \frac{\partional \, \ordial \cJ}{\partional \, \ordial \bfQ}\ordial \cdot \bfQ  \es  \mbox{ constant }  \m .
\label{en-int-2}
\ee
Suppose further that the equations corresponding to 1)
$$
0  \es  \frac{\partional \, \ordial \cJ}{\partional \bfm}(\bar{\bfQ}, \ordial \bar{\bfQ}, \bfm) \m \mbox{ can be solved for  } \m \bfm  \m .
$$
One can then pass from 
\be 
\ordial \cJ (\bar{\bfQ}, \ordial \bar{\bfQ}, \bfm)
\ee 
to a reduced 
\be 
\ordial \cJ_{\sr\se\sd}(\bar{\bfQ}, \ordial\bar{\bfQ})  \m :
\ee 
{\it multiplier elimination}.

\m 

\n Configuration--change space and configuration--velocity space are conceptually distinct presentations of the same tangent bundle $\FrT(\FrQ)$.  
Formulation in terms of change $\ordial \bfQ$ can furthermore be viewed as introducing a {\it change covector}.
This is in the sense of inducing `{\it change weights}' to Principles of Dynamics entities, 
analogously to how introducing a conformal factor attaches conformal weights to tensors.
For instance, $\ordial\fs$ and $\ordial\cJ$ are change covectors as well.
On the other hand, ${\cal S}$ is a change scalar: an entity which remains invariant under passing from the standard Principles of Dynamics to TRiPoD, 
due to their being already-TRi.

\m 

\n TRiPoD's formulation of momentum is  
\be
\bfP  \:=  \frac{\partional \, \ordial \cJ}{\partional \, \ordial \bfQ}  \m ,  
\label{TRi-Mom-2}
\ee
which is a change scalar as well.

\subsection{Free end notion of space variation}\label{FENoS}

Suppose a formulation's multiplier coordinate $\fm$ is replaced by a cyclic velocity $\fc$ \cite{ABFO, FEPI} or a cyclic differential 
$\ordial\fc$ \cite{FileR}. 
The zero right hand side of the multiplier equation is replaced by $f(\mbox{notion of space alone})$ in the corresponding cyclic equation.  
However, if the quantity being replaced is an entirely physically meaningless auxiliary, 
in the cyclic formulation, the meaninglessness of its values at the end notion of space becomes nontrivial. 
I.e.\ {\it free end notion of space} variation alias {\it variation with natural boundary conditions}) \cite{CH, Fox, BrMa, Lanczos} is the appropriate procedure.  
This is a portmanteau of {\it free end point} variation for finite theories, and {\it free end spatial hypersurface} variation for Field Theories.\footnote{To be 
clear, `free end' here refers to free value {\sl at} the end notion of space rather than the also quite commonly encountered freedom {\sl of} the end notion of space itself.} 
Such a variation imposes more conditions than the more usual fixed-end variation does: three conditions per variation, 
\beq
\frac{\partional \, \ordial\cJ}{\partional\bfg} = \ordial{\bfp}^{\sfg} \mma \mbox{alongside } \m  
\left.
\bfp^{\sfg}
\right|_{\mbox{\scriptsize end}} = 0 \m .  
\label{correct}
\eeq
Case 1) If the auxiliaries $\bfg$ are multipliers $\bfm$, (\ref{correct}) just reduces to
$$
\bfp^{\sfg} \es 0  \mma  \frac{\partional\cJ}{\partional\bfm}  \es  0
$$ 
and redundant equations.
So in this case, the end notion of space terms automatically vanish by applying the multiplier equation to the first factor of each.  
This holds regardless of whether the multiplier is not auxiliary and thus standardly varied, or auxiliary and thus free end notion of space varied. 
This is because this difference in status merely translates to whether or not the cofactors of the above zero factors are themselves zero.  
Consequently the free end notion of space subtlety in no way affects the outcome in the multiplier coordinate case.  
This probably accounts for the above subtlety long remaining unnoticed. 

\m 

\n Case 2) If the auxiliaries $\bfg$ are considered to be cyclic coordinates $\bfc$, (\ref{correct}) reduces to 
\beq 
\left.
\bfp^{\sfg}
\right|_{\se\sn\sd-\sN\sO\sS}  \es  0 \m 
\label{ckill}
\eeq
alongside
$$
\dot{\bfp}^{\sfg}    = 0 \m  \mbox{(or equivalently} \m 
\ordial{\bfp}^{\sfg} = 0) 
$$
\beq
\m \Rightarrow \m \m
\bfp^{\sfg} = \bC(\mbox{notion of space}) \mbox{ , invariant along the curve of notion of space} \m . 
\label{hex}
\eeq 
$\bC(\mbox{notion of space})$ is now identified as $0$ at either of the two end notion of space (\ref{ckill}).
Since this is invariant along the curve of notions of space, it is therefore zero everywhere.  
So (\ref{hex}) and the definition of momentum give 
$$
\frac{\partional \cL}{\partional \dot{\bfc}}            \:=  \bfp^{\sfg} \m  \mbox{ or equivalently }  \m
\frac{\partional \ordial\cJ}{\partional \ordial{\bfc}}  \es  0                                         \m .  
$$
In conclusion, the above free end point notion of space working ensures that the cyclic and multiplier formulations of auxiliaries in fact give {\sl the same} variational equation. 
Thus complying with Temporal Relationalism by passing from encoding one's {\sl auxiliaries} as multipliers 
to encoding them as cyclic velocities or differentials is valid without spoiling the familiar and valid physical equations.

\m 

\n{\bf Remark 1} A similar working \cite{FEPI} establishes that passage to the Routhian for an auxiliary formulated in cyclic terms reproduces the outcome of   
                                                         multiplier elimination for that same auxiliary formulated in terms of multipliers.

\subsection{TRi Legendre transformation}

One can now apply Legendre transformations that inter-convert changes $\ordial \bfQ$ and momenta $\bfP$. 

\m 

\n Example 1) {\it Passage to the $\ordial$-Routhian}  
\beq
\ordial \cR(\bar{\bfQ}, \ordial \bar{\bfQ}, \bfP^{\scc}) \:=  \ordial \cJ(\bar{\bfQ}, \ordial \bar{\bfQ}, \ordial \bfc) - \bfP^{\scc} \cdot \ordial \bfc  \m .
\label{d-Routh}
\eeq
passage to the $\ordial$-{\it Routhian} furthermore requires being able to 
$$ 
\mbox{solve  } \m const_{\sfY}  \es  \frac{\partional \, \ordial \cJ}{\partional \, \ordial \bfc}(\bar{\bfQ}, \ordial \bar{\bfQ}, \ordial \bfc)
\m \mbox{ as equations for the } \m \ordial \bfc \m .
$$
This is followed by substitution into (\ref{d-Routh}).
One application of this is the passage from Euler--Lagrange type actions to the geometrical form of the Jacobi actions, now done without ever introducing a parameter; 
another is Article I's reduction procedure.  

\m 

\n{\bf Example 2)} {\it Passage to the $\ordial$-anti-Routhian}, 
\beq
\ordial \cA(\bar{\bfQ}, \bar{\bfP}, \ordial \bfc)  \es  \ordial \cJ(\bar{\bfQ}, \ordial \bar{\bfQ}, \ordial \bfc) - \bfP^{\scc} \ordial \bar{\bfQ}  \m .
\eeq
A subcase of this plays a significant role in the next Section.

\m 

\n{\bf Example 3)} {\it Passage to the $\ordial$-Hamiltonian}, 
\beq
\ordial \cH(\bfQ, \bfP)  \es  \bfP \cdot \ordial \bfQ - \ordial \cJ(\bfQ, \ordial \bfQ)  \m .  
\eeq
The corresponding equations of motion are in this case $\ordial${\it -Hamilton's equations} 
\beq
\frac{\partional \, \ordial \cH}{\partional \bfP}  \es    \ordial \bfQ  \mma  
\frac{\partional \, \ordial \cH}{\partional \bfQ}  \es  - \ordial \bfP  \m . 
\eeq

\subsection{TRi-morphisms}

\n{\bf Structure 1} Suppose we are to keep no cyclic differentials.
The usual $\FrQ$ morphisms apply, except that specifically $Point$ rather than $Point_t$ is involved.
Temporal Relationalism also requires use of $Can$ rather than $Can_t$ in the $\ordial$-Hamiltonian formulation.  

\m 

\n{\bf Structure 2} Some cases here involve augmenting $\Abs$ to 
\be 
\Abs \times \Int
\ee 
for $\Int$ an internal space. 

\m 

\n{\bf Remark 1} Configurational Relationalism has hitherto in Part II rested on Mach's Space Principle.  
To continue to have such a supporting element, we now need to paraphrase a `Mach-type Internal Principle' to accompany it.  
`No one is competent to predicate things about gauge-dependent properties of internal space or motion thereover. 
These are pure things of thought, pure mental constructs that cannot be produced in experience. 
All our principles of Gauge Theory are, as we have shown in detail, experimental knowledge concerning gauge-independent quantities'.

\m 

\n{\bf Structure 3}  Let us next consider 
\be 
Aut\big(\Abs \times \Int\big) = Aut\big(\Abs\big) \times Aut\big(\Int\big) \m ,
\ee 
or some subgroup 
\beq
 \lFrg_{\se\sx\st} \times \lFrg_{\si\sn\st} \m 
\label{ext-x-int}
\eeq
of this, where `ext' standing for external transformations and `int' for internal ones (in the same sense as in Particle Physics).

\m

\n{\bf Structure 4}  We use $\lFrg$-correcting cyclic ordial (ordinary or partial) differential portmanteau auxiliaries $\ordial\bfg$.  
Encoding one's $\lFrg$ auxiliary variables in either of the above ways continues to require subsequent care with how one performs one's Calculus of Variations.
In the portmanteau case, this entails the free end notion of space variational portmanteau.  

\m 

\n{\bf Structure 5}  The corresponding action is 
\beq
{\cal S}  \es  \iint_{\sN\so\sS} \ordial \mN\mo\mS \, \ordial \cJ  
     \es  \iint_{\sN\so\sS} \, \ordial \mN\mo\mS \sqrt{2} \ordial \fs \sqrt{\cW}   \m , 
\label{S-Port}
\eeq
\beq
\ordial \fs           \:=  ||\ordial_{\sbfg} \bfQ ||_{\sbfM}  \m \mbox{ and } \m 
\ordial_{\sbfg} \bfQ  \:=  \ordial{\bfQ} - \stackrel{\rightarrow}{\lFrg}_{\sordial{\sbfg}} \bfQ                                 \m .   
\eeq 
A field-theoretic update of Article V's table of formulations is as follows.  

\begin{tabbing}

\underline{\bf \scriptsize Type of variables}               \=             
\underline{\scriptsize for which the}                       \= 
\underline{\bf \scriptsize key portmanteau}                 \=  
\underline{\scriptsize gives the}                           \= 
\underline{\bf \scriptsize equations of motion }            \\

{\scriptsize Lagrangian}        \m                          \>           
{\scriptsize $\sbfQ, \frac{\sordial{\sbfQ}}{\sordial t}$}   \>
{\scriptsize Lagrangian}                                    \>
{\scriptsize  $\cL$}                                        \>
{\scriptsize Euler--Lagrange}                               \\

{\scriptsize Machian}          \m                           \>                
{\scriptsize $\sbfQ, \sordial {\sbfQ}$}                     \>     
{\scriptsize Jacobi arc element}                            \> 
{\scriptsize $\sordial\cJ$}                                 \> 
{\scriptsize `Jacobi--Mach'}                                \\

{\scriptsize Hamiltonian}      \m                           \>
{\scriptsize $\sbfQ, \sbP$}                                 \>                      
{\scriptsize Hamiltonian}                                   \> 
{\scriptsize $\cH$}                                         \> 
{\scriptsize Hamilton's}                                    \\

\end{tabbing}

\n{\bf Remark 2} The Hamiltonian formulation is unadulterated both by passing from Lagrangian to Machian variables and by bringing in portmanteau derivatives.  
Upon including the $\lFrg$ auxiliaries, however, there is a slight alteration to the Hamiltonian formulation. 
This is from the usual `extended' Hamiltonian (Article II) to the {\it `extended' $\ordial$A-Hamiltonian}; 
throughout this Series of Articles, A- stands for `almost'.
This is however just reformulating the {\sl unphysical} sector of the theory. 
A second table now also incorporating this expansion is as follows.  

\begin{tabbing}

\underline{\bf \scriptsize Type of variables}        \hspace{0.08in}              \= 
\underline{\scriptsize for which the}                \hspace{0.08in}              \=  
\underline{\bf \scriptsize key portmanteau}          \hspace{0.08in}              \=  
 \hspace{0.27in} \underline{\scriptsize gives the}   \hspace{0.08in}              \= 
\underline{\bf \scriptsize equations of motion}                                   \\

{\scriptsize Lagrangian}                                                          \> 
$\sbfQ,\sbfm,  \frac{\sordial \sbfQ}{\sordial t}$                                 \>
{\scriptsize Lagrangian}                                                          \>
{\scriptsize  $\cL$}                                                              \>
\hspace{0.3in} {\scriptsize Euler--Lagrange}                                      \\

{\scriptsize Machian}                                                             \> 
$\sbfQ, \sordial \sbfQ, \sordial \bsfc$                                           \>     
{\scriptsize Jacobi arc element}                                                  \> 
$\sordial${\scriptsize $\cJ$}                                                     \> 
\hspace{0.3in} {\scriptsize `Jacobi--Mach'}                                       \\

{\scriptsize total Hamiltonian}                                                   \> 
{\scriptsize $\sbfQ, \bfm, \sbfP$}                                                \>                      
{\scriptsize total Hamiltonian}                                                   \> 
{\scriptsize $\cH_{\tT\to\ttt\ta\tl} = \cH + \bfm \cdot \btcF$}                   \> 
\hspace{0.3in} {\scriptsize Hamilton's}                                           \\

{\scriptsize $\sordial$A-total Hamiltonian }                                      \> 
{\scriptsize $\sbfQ, \sbfP, \sordial\bfc$ }                                       \>                      
{\scriptsize $\sordial$A-total Hamiltonian}                                       \> 
{\scriptsize $\sordial\cH_{\tT\to\ttt\ta\tl} 
= \sordial\cH + \sordial\bfc \cdot \btcF$}                                        \> 
\hspace{0.3in} {\scriptsize $\sordial$A-Hamilton's}                               \\

\end{tabbing}

\n{\bf Structure 6}  The conjugate momenta are then (using the {\it partional derivative} portmanteau of partial and functional derivatives)  
\be
\bfP        \:=  \partional \frac{\ordial\cJ}{\partional \, \ordial{\bfQ}}   
            \es  \u{\u{\bfM}} \sqrt{\frac{\cW}{\fT}} \, \ordial_{\sbfg} \u{\bfQ} \m .  
\ee
\n{\bf Structure 7} These obey one primary constraint per relevant notion of space point, interpreted as an equation of time,  
\beq
\Chronos  \:=  \half {||\biP||_{\sbiN}}^2 - \cW( \biQ)  
          \es  0                                                  \m . 
\eeq
Thus it is purely quadratic in the momenta.

\m 

\n{\bf Structure 8} The $\bfP$ also obey some secondary constraints per relevant notion of space point from variation with respect to $\bfg$.  
\beq
0  \es  \partional \frac{  \ordial \cJ  }{  \partional \, \ordial\bfc^{\sfg}  } 
   \:=  \bShuffle 
   \es 
\frac{  \mbox{\Large $\delta$} \{\stackrel{\rightarrow}{\lFrg_{\sordial\sfc}}\fQ^{\sfA}\}  }{  \mbox{\Large $\delta$} \ordial\bfc^{\sfg}  }  \fP_{\sfA} \m ;     
\label{LinL}
\eeq
these are linear in the momenta, and so are also denoted by (for now) $\bLin$.  

\m 

\n{\bf Structure 9}  Let us next denote the joint set of these constraints by $\bscF$, 
under the presumption that they are confirmed to be first-class in Articles III and VII.  
The indexing set designation assumes there is only one quadratic constraint, so all our examples' $\fF$ ranges over $\fG$ and the one quadratic value.   

\m 

\n{\bf Structure 10}  The corresponding Jacobi--Mach equations of motion are  
\beq
\ordial   \frac{ \partional \,\ordial\cJ  }{  \partional \,\ordial \bfQ^{\sfA}  } \es  \frac{  \partional \, \ordial\cJ  }{  \partional \fQ^{\sfA}  }    \m \m \Rightarrow \m \m
\eeq
\beq
\frac{\sqrt{2 \cW}}{  ||\ordial \bfQ||_{\sbfM}  } \ordial  
\left\{
\frac{\sqrt{2 \cW}}{||\d \bfQ||_{\sbfM}} \ordial \fQ^{\sfA}
\right\} 
\m + \m \Gamma^{\sfA}\mbox{}_{\sB\sfC} \frac{\sqrt{2 \cW}}{|| \ordial  \biQ||_{\sbfM}} \ordial \fQ^{\sfB} 
  \frac{\sqrt{2 \cW}}{|| \ordial  \biQ||_{\sbfM}} \ordial \fQ^{\sfC}  \es  \fN^{\sfA\sfB}\frac{\partional \fW}{\partional \fQ^{\sfB}}                    \m .
\label{Evol-2}
\eeq
The previous Section's Best Matching procedure admits the following generalization. 

\m

\n{\bf TRi Best Matching 0)} Start with the `arbitrary $\lFrg$ frame corrected' action (\ref{S-Port}).  

\m 

\n{\bf TRi Best Matching 1)} Extremize over $\lFrg$.
This produces a constraint equation $\bShuffle$ that is of the form $\bLin$: linear in the momenta.   

\m 

\n{\bf TRi Best Matching 2)} The Machian variables form of this equation, with Machian data $(\bfQ, \ordial \bfQ)$ is to be solved for the $\ordial\bfg$ themselves.

\m 

\n{\bf TRi Best Matching 3)} Substitute this solution back into the action: an example of passage to the $\ordial$-Routhian (see Appendix of VII).  
Again this produces a final $\lFrg$-independent expression that could have been directly arrived at as a {\sl direct} implementation of Configurational Relationalism ii). 

\m 

\n{\bf TRi Best Matching 4)} Finally elevate this new action to be one's new starting point.  

\m 

\n{\bf TRi Best Matching 3$^{\prime}$)} As a distinct application of TRi Best Matching 2), emergent Machian times are now of the general form 
\beq  
\ft^{\se\sm} = \cF \lfloor \bfQ, \ordial \bfQ \rfloor \m , 
\eeq
a particular realization of which is $\ft^{\se\sm}$ of e.g. the Jacobi type,
\beq
\lft^{\se\sm}_{\sR\si}  \es  \lE^{\prime}_{\sbfg \, \in \,  \nFrg}    \int \frac{||\ordial_{\sbfg}\bfQ||_{\sbfM}}{\sqrt{2 \, \cW}}  \m . 
\label{Kronos2}
\eeq
If one succeeds in carrying out Best Matching, moreover, both the two-functional and implicit-formulation complications are washed away.
This leaves $\lft^{\se\sm}_{\sR\si}$ expressed in terms of the {\sl reduced} $\FrQ$'s geometry.

\m 

\n{\bf Remark 3} The above expression does {\sl not} contain a spatial integral: the field-theoretic $\lt^{\se\sm}$ is {\sl local}.  
Moreover, the essential line of thought of this Section is the {\sl only} known approach to Configurational Relationalism 
that is general enough to cover the Einstein--Standard Model presentation of Physics. 

\m 

\n{\bf Structure 1} The momenta in terms of the corresponding derivative $\Last$ are 
\beq
\u{\bfP} = \u{\u{\bfM}} \cdot \Last \u{\bfQ}  \m .  
\eeq
\n{\bf Structure 2} The equations of motion now take the `parageodesic' form 
\beq
\Last\Last  \fQ^{\sfA} + \Gamma^{\sfA}\mbox{}_{\sfB\sfC}\Last \fQ^{\sfB}\Last \fQ^{\sfC}  \es  \fN^{\sfA\sfB} \frac{\partional \cW}{\partional \fQ^{\sfB}}  \m .
\label{parag-2} 
\eeq 
It can also be cast as a true `geodesic' equation with respect to the physical metric whose line element is $\ordial\cJ$. 
Finally one of the evolution equations per relevant notion of space point can be supplanted by 
the emergent Lagrangian form of the quadratic `energy-type' constraint (I.98).

\section{Example 1) Electromagnetism by itself}\label{EM-Rel}

\n{\bf Structure 1} Consider the space of 1-forms on $\mathbb{R}^3$ in the role of $\FrQ$.    

\m 

\n{\bf Structure 2} $\lFrg = Diff(\mathbb{R}^3)$ is not applied to flat-space Electromagnetism because $\delta_{ij}$ breaks this in the active sense.   
However, $\lFrg = Rot(3)$ can be considered. 
Internal Relationalism involving $U(1)$ clearly also applies. 

\m 

\n{\bf Remark 1} The latter works out fine for this (including using a $\dot{\Psi}$ or $\pa\Psi$ auxiliary in place of the electric potential $\Phi$ \cite{FEPI}). 
This gives in each case the expected Gauss constraint $\scG$.  
However Spatial and Temporal Relationalism is prohibitively restrictive in the case of Electromagnetism.  
E.g.\ involving $Diff(\bupSigma)$ gives that \cite{Lan2} the Poynting vector must vanish: 
\be 
\underline{\mE} \cr \underline{\mB} = 0 
\ee
I.e.\ 
\be 
\underline{\mE} = 0 \mma 
\underline{\mB} = 0 \m \mbox{ or } \m 
\underline{\mE} \parallel \underline{\mB} \mbox{ (killing signal propagation)} \m .
\ee
In any case, Electromagnetism by itself has background structures (typically the Minkowski metric $\boldeta$, 
                                                                   or the Euclidean metric $\bdelta$ on flat spatial slices).  

\m 

\n{\bf Remark 2} The resolution of these issues is that inclusion of GR to make the Einstein--Maxwell system 
frees one from these background structures and the above zero Poynting vector restriction (see Sec \ref{GR+Fund}). 
Yang--Mills Theory and the various associated Gauge Theories follow suite in these regards.
More generally still, Field Theories of matter are found to not be properly supported in the absence of GR as regards attaining Background Independence.

\section{Example 2) GR}\label{GR-G-Q}

\n{\bf Structure 1} For this particularly substantial example, $\FrQ = \Riem(\bupSigma)$ -- 
the space of Riemannian 3-metrics on some fixed spatial topological manifold $\bupSigma$ 
that is taken to be compact without boundary\ both for simplicity and for Machian reasons. 
Moreover, equipping $\bupSigma$ with $\bh$ requires a more involved form than RPMs' multiple copies of absolute space; 
indeed GR gives further reason to adopt Sec II.5.2's group action on configuration space rather than on space.

\m 
 
\n{\bf Structure 2} We next need to deal with the Configurational Relationalism so far breaking the Manifest Reparametrization Invariance.
The group of physically irrelevant motions $\lFrg$ is usually taken to be $Diff(\bupSigma)$: the diffeomorphisms on $\bupSigma$; 
see Article IX for further alternatives.
\n A further example of structural compatibility between $\FrQ$ and $\lFrg$ that manifests itself in Geometrodynamics 
is $Diff$ being based upon the same underlying topological manifold $\bupSigma$ that $\Riem$ is.

\subsection{Baierlein--Sharp--Wheeler action}\label{BSW-San}

Let us first consider the Baierlein--Sharp--Wheeler (BSW) action ${\cal S}_{\sB\sS\sW}$ (II.31) \cite{BSW}.
Because $\lambda$ and $t$ coincide for GR due to its status as an already-parametrized theory, 
the distinction between the BSW kinetic term and the ADM one is entirely conceptual rather than mathematical.  
The equivalence of the ADM and BSW actions for GR is then established by the multiplier elimination move done for Minisuperspace in Article I  
immediately carrying over to GR in general \cite{BSW, BY1}.

\m 

\n GR's configuration space metric is furthermore indeed built out of the dynamical variables: 
\be 
\mM^{ijkl} = \sqrt{\mh}\{\mh^{ik}\mh^{jl} - \mh^{ij}\mh^{kl}\}   \m , 
\ee 
which successfully implements Configurational Relationalism Postulate i).  

\m 

\n The Thin Sandwich formulation \cite{BSW, WheelerGRT} consists of the following. 

\m 

\n{\bf Thin Sandwich 0)} Consider the BSW action.  

\m 

\n{\bf Thin Sandwich 1)} Vary this to obtain the constraint equation $\u{\bscM}$.  

\m 

\n{\bf Thin Sandwich 2)} Consider the `Thin Sandwich equation', i.e.\ the Lagrangian-variables form of $\u{\bscM}$, 
\be 
{\cal D}_{j}  \left\{ \frac{  \sqrt{  {\cal R} - 2 \, \slLambda  }  }
                           {  ||\delta_{\vec{\beta}} \bh ||_{\sbM}  }    
                          \{ \mh^{jk} {\delta^{l}}_{i} - {\delta^{j}}_{i} \mh^{kl} \}\{ \delta_{\vec{\upbeta}}   \mh_{kl} \}  \right\}  \es  0                               \m . 
\ee
or, as an explicit PDE in components, 
\be 
{\cal D}_{j}\left\{ {\sqrt{\frac{{\cal R} - 2 \, \slLambda}
{\{\mh^{ac}\mh^{bd} - \mh^{ab}\mh^{cd}\}\{\pa{\mh}_{ab} - 2{\cal D}_{(a}\upbeta_{b)}\}\{\pa{\mh}_{cd} - 2{\cal D}_{(c}\upbeta_{d)}\}  }}} 
 \{\mh^{jk} \delta^{l}_{i} - \delta^{j}_{i}\mh^{kl}\}\{\pa{\mh}_{kl} - 2{\cal D}_{(k}\upbeta_{l)}\}\right\} = 0 \m .   
\label{Thin-San-Eq}
\eeq
Alongside `thin sandwich data' 
\be
(\bh, \dot{\bh})  \m , 
\label{Thin-San-Data}
\ee
this constitutes a PDE problem (equations plus data) to be solved for the shift $\u{\upbeta}$: the Thin Sandwich Problem. 

\m 

\n{\bf Thin Sandwich 3.a)}    Construct 
\be 
\pounds_{\underline{\upbeta}} \bh                                                              \m :
\ee 
which is a $\lFrg$-act realized as  
\be 
\vec{Diff}_{\underline{\upbeta}} \bh                                                           \m .
\ee  
Then   
\be 
\updelta_{\vec{\upbeta}}\bh  =  \dot{\bh} - \pounds_{\underline{\upbeta}} \bh                  \m .
\ee 
\n{\bf Thin Sandwich 4.a)} Next construct an emergent counterpart to $\upalpha$, 
\be 
\mN := \sqrt{  \frac{  \mT^{\sG\sR}_{\sB\sS\sW}  }{  4\{ {\cal R} - 2 \, \slLambda \} }  }              \m .
\ee 
\n{\bf Thin Sandwich 5)} Thin Sandwich 3.a) and 4.a) permit one to construct the extrinsic curvature 
\be 
{\bcalK}  \es  {\bcalK}(\underline{x}; \bh, \underline{\upbeta}, \mN]
\ee  
using the computational formula
\beq
{\bcalK}  \es  \frac{\updelta_{\vec{\upbeta}}{\mh}_{ij}}{2 \, \mN}                             \m , 
\label{Kij-BSW}
\eeq 
which is the last form in (II.11) except that BSW's emergent $\mN$ has taken the place of ADM's presupposed $\upalpha$.

\subsection{The Thin Sandwich Problem}

Unfortunately the Thin Sandwich Problem, (\ref{Thin-San-Eq}, \ref{Thin-San-Data}) is hard to handle as a PDE problem.  
See Appendix O of \cite{ABook} for an outline of existence and uniqueness results for this, 
for which \cite{BO69, BF, FodorTh} constitute original literature by Belasco, Ohanian, Bartnik and Fodor.    
Generic GR solution of this equation is, moreover, out of the question.  
Since the Thin Sandwich equation has a square root trapped inside the ${\cal D}_i$, a fairly complicated PDE ensues.

\m 

\n{\bf Easier examples} Contrast how in RPM (Articles II, V) -- and even in the SIC case (Article XI) to leading order -- 
Best Matching gives a merely algebraic equation which is much easier to handle.\footnote{In the RPM case, this at least holds locally, 
due to zeros if the moment of inertia being encountered in collinear configurations.}    

\m 

\n The full GR case being harder is rooted in $\lFrg = Diff(\bupSigma)$.

\subsection{Reparametrization-Invariant relational action for GR}\label{MRI-GR}

The BSW action does succeed in being formulated free from a extraneous background time-like notion such as the GR lapse.  
However, this does not comply with Temporal Relationalism since the presence of the shift $\upbeta^i$ breaks Manifest Reparametrization Invariance.
None the less, Article I has laid out how to get round this deficiency. 

\m 

\n To link between the two formalisms, we first introduce the {\it cyclic velocity of the frame} (Fig \ref{Strutty}.a)   
\be 
\dot{\u{F}}  = \frac{\pa \u{F}}{\pa \lambda}
\ee
This is computationally equal to the shift $\upbeta^i$.
We can then form the `BFO-A' action \cite{RWR, ABFO}, whose conceptual form is  
\beq
{\cal S}_{\sM\sR\sI}^{\sG\sR}      \es  \int \d\lambda \int_{\sbupSigma}\sqrt{\overline{\cal T}_{\sM\sR\sI}^{\sG\sR}   \overline{{\cal R} - 2 \, \slLambda}} 
\mma 
\overline{cal T}_{\sM\sR\sI}^{\sG\sR}  \:=  {\left|\left|  \Circ_{\sF}  \right|\right|_{\sbM}}^{2}
                                   \es  \left|\left|  \dot{\bh} - \pounds_{{\dot{\u{\sF}}}}\bh  \right|\right|_{\sbM}^{\m\m 2}                     \m . 
\label{S-relational} 
\eeq
This implements both Temporal and Configurational Relationalisms.

\m 

\n Such replacement of shift by velocity of the frame can be done for the ADM action as well. 
If accompanied homogeneously by replacing the lapse by the velocity of the instant, $\mI$ (Fig \ref{Strutty}.a), we obtain an `A action' \cite{FEPI} 
analogue of the ADM action. 

\m 

\n{\bf Remark 1} Passing from \cite{FEPI, ABook} the A action for GR    to the BFO-A action for GR is passage to the Routhian \cite{Lanczos} 
in exact parallel to that from Euler--Lagrange to Jacobi actions for Mechanics \cite{Lanczos, Arnold}. 
%
{            \begin{figure}[!ht]
\centering
\includegraphics[width=0.47\textwidth]{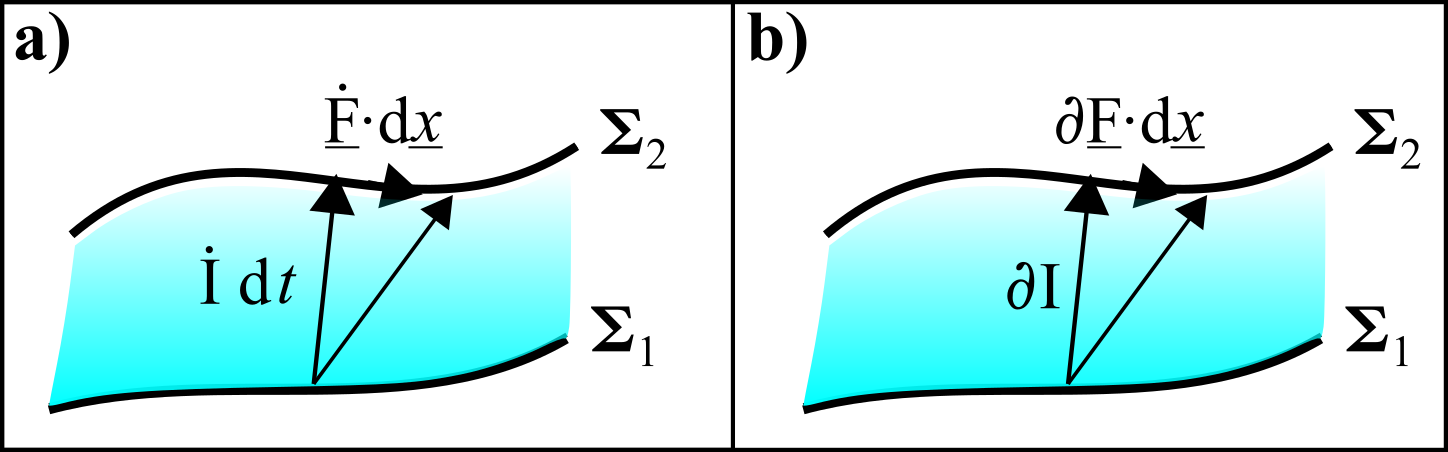}
\caption[Text der im Bilderverzeichnis auftaucht]{        \footnotesize{TRi upgrade of Fig II.1's ADM shift and lapse split. 

\m

\n a) velocity of the frame and velocity of the instant split of spacetime. 

\m 

\n b) Differential of the frame and differential of the instant split of spacetime.} }
\label{Strutty}\end{figure}            }

\subsection{Geometrical action for GR}\label{S-GR-Geom}

The Author subsequently gave a timeless $\d\u{F}$-corrected version of BSW-type action for GR \cite{PPSCT}; 
see \cite{FileR, ABook} for details, and Fig \ref{Strutty}.b) for details of the corresponding spacetime split (when meaningful).
This is of Manifestly Parametrization Irrelevant type, or, dually of geometrical type; let us refer to it as the {\it relational action for GR}. 
Now using the {\it cyclic partial differential of the frame} auxiliary, $\pa \mF^i$, the final Relationalism-implementing action for GR is 
\beq
{\cal S}^{\sG\sR}_{\sR\se\sll}  \es  \iint_{\sbSigma}\d^{3}x \, \pa{\cal J}  
                   \es  \iint_{\sbSigma}\d^{3}x     \sqrt{\overline{{\cal R} - 2 \, \slLambda}} \, \pa\ms^{\sG\sR}_{\sR\se\sll}
\label{S-Rel-d-2}
\eeq
\beq
\mbox{for } \m \pa\ms^{\sG\sR}_{\sR\se\sll}  \:=  ||\pa_{\usF}\bh||_{\sbM}  \m \mbox{ and }      \m 
\pa_{\usF}\mh_{ij}              \:=    \pa{\mh}_{ij} - \pounds_{\pa{\usF}}\mh_{ij}  \m .
\label{T-Rel-d}   
\eeq

\subsection{How the geometrical action for GR works}\label{S-GR-Geom-W}

{\bf Structure 1} The conjugate momenta are 
\beq
\u{\u{\bp}}  \:=  \frac{  \updelta \, \pa {\cal J}  }{  \updelta \, \pa \, \u{\u{\bh}}  }  
          \es   2 \sqrt{  \overline{{\cal R} - 2 \, \slLambda}  } \, \u{\u{\u{\u{\bM}}}} \frac{  \pa_{\usF}\u{\u{\bh}}  }{  \pa \ms^{\sG\sR}_{\sR\se\sll}  }  \m . 
\label{tumvel}
\eeq
\n{\bf Structure 2} The GR Hamiltonian constraint $\scH$  now follows as a primary constraint that is purely quadratic in the momenta.  

\m 

\n{\bf Structure 3} The GR momentum constraint $\u{\bscM}$ arises as a secondary constraint from variation with respect to the auxiliary $Diff(\bupSigma)$-variables $\u{\mF}$; it is linear in the momenta.  

\m 

\n{\bf Structure 4} The Jacobi--Mach equations of motion are (using $\slLambda = 0$ for simplicity)   
\beq
2 \sqrt{ \overline{{\cal R}} } \, \frac{\pa_{  \underline{\sF}}\u{\u{\bp}}  }{  \pa \ms^{\sG\sR}_{\sR\se\sll}  }  \es 
\left\{ 
\sqrt{\mh} \big\{ {\cal R}    \, \u{\u{\bh}} - \, \u{\u{\bcalR}} + \u{\bcalD} \, \u{\bcalD} - \u{\u{\bh}} \, \triangle  \big\}  
- \mbox{$\frac{  2  }{  \sqrt{\sh}  }$} \left\{ \u{\u{\bp}} \cdot \u{\u{\bp}} - \frac{\mp}{2} \, \u{\u{\bp}} \right\} 
\right\} 
\frac{\pa\ms^{\sG\sR}_{\sR\se\sll}}{2\sqrt{\overline{{\cal R}}}}  \m .  
\label{BFO-Evol2}
\eeq
Via the Bianchi identity \cite{Wald}, these immediately propagate the above constraints without producing further conditions.

\subsection{TRi form of the Thin Sandwich} 

\n Let us next reiterate the Thin Sandwich procedure in the TRi formulation's manifestly temporally Machian form.  

\m

\n{\bf TRi Thin Sandwich 0)} Consider the relational GR action (\ref{S-Rel-d-2}) \cite{Lan2, AM13}.  

\m 

\n{\bf TRi Thin Sandwich 1)} Vary it with respect to $\u{\mF}$ to obtain the constraint equation $\u{\bscM}$ \cite{ABFO}.  

\m 

\n{\bf TRi Thin Sandwich 2)} Consider the `Machian Thin Sandwich equation': the Machian-variables form of $\u{\bscM}$,
\be 
{\cal D}_{j}  \left\{ \frac{  \sqrt{  {\cal R} - 2 \, \slLambda  }  }
                           {  ||\Circ_{\pa\u{\sF}} \bh ||_{\sbM}  }    
                          \{ \mh^{jk} {\delta^{l}}_{i} - {\delta^{j}}_{i} \mh^{kl} \}\{ \delta_{\pa{\u{\sF}}} \mh_{kl} \}  \right\}  \es  0                               \m . 
\ee
or, as an explicit PDE in components, 
\be 
{\cal D}_{j}  
\left\{ 
\sqrt{  \frac{  {\cal R} - 2 \, \slLambda  }{  \{\mh^{ac}\mh^{bd} - \mh^{ab}\mh^{cd}\}\{\pa{\mh}_{ab} - 2{\cal D}_{(a}\pa\mF_{b)}\}\{\pa{\mh}_{cd} - 2{\cal D}_{(c}\pa\mF_{d)}\}  }    }
                                               \{\mh^{jk} \delta^{l}_{i} - \delta^{j}_{i}\mh^{kl}\}\{\pa{\mh}_{kl} - 2{\cal D}_{(k}\pa{\mF}_{l)}\}\right\} = 0 \m ,
\label{TRi-Thin-San-Eq}
\eeq
with `TRi thin sandwich data'  
\beq
(\bh, \pa {\bh})  \m ,  
\label{TRi-Thin-San-Data}
\eeq 
as an equation for the partial differential of the frame auxiliary $\pa\u{\mF}$.  
Altering (\ref{Thin-San-Eq}, \ref{Thin-San-Data}) to (\ref{TRi-Thin-San-Eq}, \ref{TRi-Thin-San-Data}), moreover, 
makes no difference to the {\sl mathematical form} of this PDE problem.  

\m 

\n{\bf TRi Thin Sandwich 3.a)} Construct $\pounds_{\pa\underline{\sF}} \bh$ and then the Best Matching corrected derivative   
\beq
\pa_{\underline{\sF}}\bh  =  \pa \bh - \pounds_{\pa\underline{\sF}} \bh    \m . 
\label{BM-Der}
\eeq
This is a distinct conceptualization of the same mathematical object as the hypersurface derivative.  

\m  

\n{\bf TRi Thin Sandwich 4.a)} Construct the {\it emergent differential of the instant} 
\beq
\pa\mI  \es  \frac{  ||\pa_{\underline{\sF}} \bh||_{\sbM}  }{  2\sqrt{ \overline{{\cal R} - 2 \, \slLambda}  }  }  \m .
\label{gue}
\eeq
{\bf TRi Thin Sandwich 4$^{\prime}$)} Emergent Machian time readily follows simply from integrating up 4.a).   
4.a) moreover goes beyond BSW's own construction.
It is GR's analogue of emergent Jacobi time \cite{B94I}.  
Furthermore, it is an `all change', or, in practice `STLRC' implementation of Mach's Time Principle: 
\beq
\mt^{\se\sm}_{\sR\si}(\uix)  \es  
\mbox{\large E}^{\prime}_{\usF \, \in \,  Diff(\sbupSigma)}       
\int \frac{  ||\pa_{\usF}\bh||_{\sbM}  }{  \sqrt{     \overline{{\cal R} - 2 \, \slLambda}  }  }  \m .
\label{GRemt2}
\eeq
\n{\bf TRi Thin Sandwich 3)} Another consequence of move 2) is that one can substitute the resultant extremizing $\u{\mF}$ back into the relational GR action. 

\m 

\n{\bf TRi Thin Sandwich 4)} Take this as an ab initio new action.

\m 

\n{\bf Remark 1} Moves 1) to 4) constitute a reduction; with these, the Machian Thin Sandwich can be interpreted as a subcase of Best Matching.  

\m 

\n{\bf TRi Thin Sandwich 5)} is that moves 2) and 3) also permit construction of the extrinsic curvature through the computational formula
\beq
\u{\u{{\bcalK}}}   \es   \frac{\pa_{\underline{\sF}} \u{\u{\bh}} }{2 \, \pa \mI}         \m ,  
\label{Kij-Rel}
\eeq 
for 
\be 
{\bcalK} = {\bcalK}(\underline{x}; \bh, \pa\underline{\mF}, \pa\mI]   \m .
\ee   
This subsequently enters Spacetime Construction, as further laid out in Article IX.
See Fig \ref{Rel-BSW-Fig} for a summary so far, laying out the TRi modifications to the previous Thin Sandwich work following from the BSW action.
Also, Article XII proceeds to consider yet further completion of the thin-sandwich prescription 
in terms of constructing the whole of the universal hypersurface kinematics \cite{Kuchar76Other}.
%
{            \begin{figure}[!ht]
\centering
\includegraphics[width=0.9\textwidth]{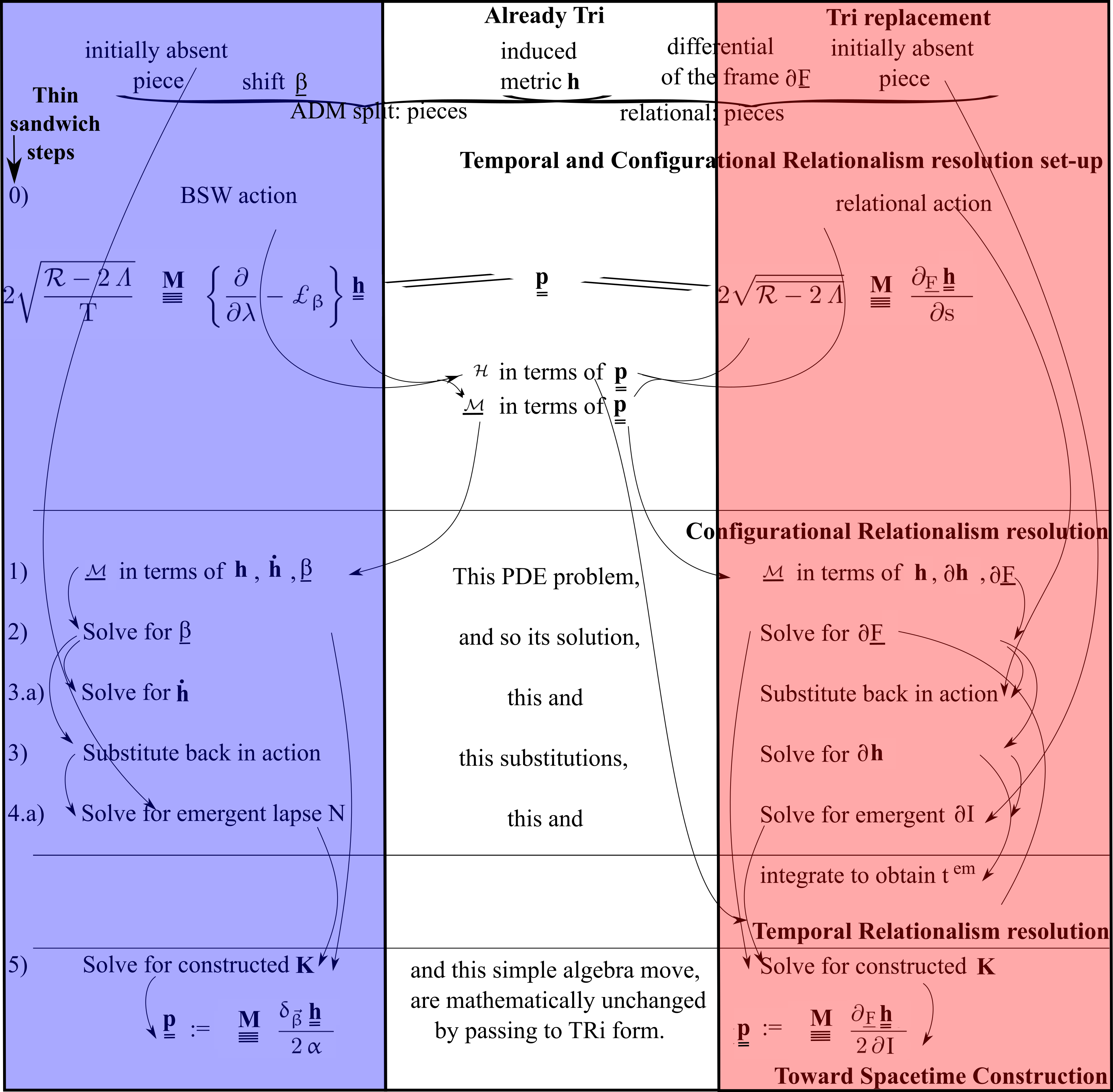}
\caption[Text der im Bilderverzeichnis auftaucht]{        \footnotesize{Standard Thin Sandwich versus TRi Thin Sandwich.   
The current Figure can be regarded as further detail of the second floor of Article XIII's TriPoD summary figure, in the case of full GR.} }
\label{Rel-BSW-Fig}\end{figure}            }

\subsection{Comments on GR's emergent Machian time}\label{t-em=t-proper}

\n{\bf Remark 1} Let us comment further on the form of expression (\ref{GRemt2}).  
It says that one has to extremize {\sl one} functional in order to use the extremal value from that in {\sl another} functional. 
This is more complicated than the usual variational problem that involves extremizing a single functional. 

\m 

\n What happens if the time functional itself is extremized?    
For finite models, this is actually equivalent to the given procedure. 
For Field Theory, however, the two become inequivalent due to factors of $\sqrt{\cW}$ 
becoming entangled with the derivative operators that arise `by parts' in the spatial integration.
The necessary property that the timestandard be independent of whether the action has already been reduced 
then forces the extremization to take the above form (Fig \ref{TEET}). 
%
{            \begin{figure}[!ht]
\centering
\includegraphics[width=0.4\textwidth]{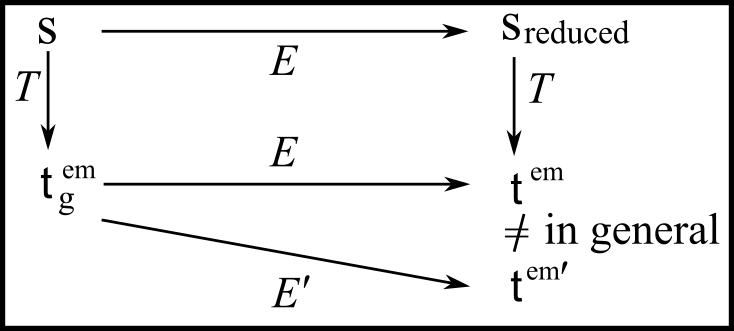}
\caption[Text der im Bilderverzeichnis auftaucht]{        \footnotesize{Let us denote the map from an action to the corresponding emergent time 
candidate by $T$,
We denote the map consisting of substituting in the $\nFrg$-extremum of the action by $E$, and the map consisting of substituting in the 
                                                    $\nFrg$-extremum of the timefunction itself by $E^{\prime}$
$E$ and $T$ naturally commute: $TE = ET$, but in general $TE^{\prime} \neq ET$. 
This is why we use the $\nFrg$-extremum of $\sFS$ in order to free t of $\fg$-dependence.} }
\label{TEET}\end{figure}            }

\m 

\n{\bf Remark 2} The local character of GR's emergent time notion arises from the field-theoretic use of local square roots 
(i.e.\ take the square root prior to integrating).  

\m 

\n{\bf Remark 3} The final form of $t^{\se\sm}$ can in this case be identified as the GR proper time, now obtained as an emergent concept.  
In a suitable cosmological setting, this is aligned with cosmic time, recovering Article I's minisuperspace result.    
We are using a distinct name and symbol $\mI$ for `instant' since this emergent entity is, most primarily, a {\it labeller of instants}. 
This ends up, very satisfactorily, being dual to the GR proper time.  
GR proper time is indeed a quantity which in general differs from point to point,  
and it is this desirable feature which arises from the field-theoretic local square root ordering (Sec I.4.6).
It is in this manner that local square roots manage to be desirable in GR despite their Finite Theory counterparts being questionably Machian 
and not physically realized.

\m 

\n Via 
$$
\pa \mI  \:=  \frac{\pa\ms}{2\sqrt{  \overline{{\cal R}}  }}  
         \es  \pa \big( \lmt^{\se\sm}_{\sR\si} \big)                 \m ,
$$ 
the $\slLambda = 0$ $\mt^{\se\sm}$--instant dual simplifies the momentum-change relation and Jacobi--Mach equations of motion
into the forms  
\be
\u{\u{\bp}}  \es  \u{\u{\u{\u{\bM}}}} \frac{\pa_{\sF}\u{\u{\bh}}}{2\,\pa\mI}  \m ,
\ee
\be
\pa_{\sF} \u{\u{\bp}}  \es  \left\{  \sqrt{\mh}     \{  {\cal R} \, \u{\u{\bh}} - \, \u{\u{\bcalR}} + \u{\bcalD} \, \u{\bcalD} - \u{\u{\bh}} \, \triangle \} 
                        - \mbox{$\frac{2}{\sqrt{h}}$} \left\{  \u{\u{\bp}} \cdot \u{\u{\bp}} - \frac{\mp}{2} \, \u{\u{\bp}} \right\}      \right\} \pa \mI  \m .  
\label{BFO-Evol3}
\ee

\section{Reduced configuration spaces for Field Theory and GR}\label{GR-Config-2}

\subsection{Topology of $\lSuperspace(\lbupSigma)$}\label{Superspace-Top}

{\bf Structure 1} Fischer showed that 
\be 
\Superspace(\bupSigma)  =  \frac{\Riem(\bupSigma)}{Diff(\bupSigma)}
\ee 
\cite{Fischer70} can be taken to possess the corresponding quotient topology.
$\Superspace(\bupSigma)$ additionally admits a metric space metric of the form \cite{Fischer70} (II.109) 
In this manner, $\Superspace(\bupSigma)$ is a metrizable topological space and thus obeys all the separation axioms and thus in particular Hausdorffness; 
it is also second-countable \cite{Fischer70}.
Superspace is thus in this way `2/3rds of a manifold'.

\m 

\n{\bf Structure 2} However, unlike $\Riem(\bupSigma)$, $\Superspace(\bupSigma)$ fails to possess the infinite-dimensional analogue 
of the locally-Euclidean property. 
Wheeler \cite{Battelle} credited Smale with first pointing this out.
Fischer \cite{Fischer70} subsequently worked out the detailed structure of $\Superspace(\bupSigma)$ as a stratified manifold.  
In particular, the appearance of nontrivial strata occurs for $\bupSigma$ that admit metrics with non-trivial $I(\langle \bupSigma, \bh \rangle)$.  
In these cases $Diff(\bupSigma)$ clearly does not act freely upon these metrics. 
Rather, the $\Superspace(\bupSigma)$ quotient space is here a stratified manifold of nested sets of strata ordered by 
$\mbox{dim}(I(\langle \bupSigma, \bh \rangle))$.\footnote{$\mbox{\sSuperspacetime}(\lFrm)$ 
also has nested strata and conical singularities corresponding to the geometries with nontrivial Killing vectors.}  
%
Indeed, Fischer \cite{Fischer70} tabulated the allowed isometry groups on various different spatial topologies.  
In this way, $\Superspace(\bupSigma)$ is not itself a manifold.

\section{Example 3) GR with fundamental matter fields}\label{GR+Fund}

In general, the Thin Sandwich equation (\ref{Thin-San-Eq}) is proportional to momentum flux $\mP_i$.  
The Thin Sandwich Problem has mostly only been considered for phenomenological matter \cite{WheelerGRT, BF, FodorTh}. 
Best Matching, however, concerns fundamental matter, a distinction of note since corrections to velocities do not occur in phenomenological matter terms.  
Giulini \cite{Giu99} did consider the Thin Sandwich Problem for Einstein--Maxwell Theory (including protective theorems).
Christodoulou \cite{Christodoulou3} and the Relational Approach \cite{RWR, AB, Van} each covered this with GR coupled to a full complement of fundamental matter.
[These works are at the level of the form of the equations but not at the level of Thin Sandwich Theorems.] 

\m 

\n The specific examples below lie within the scope of Appendix \ref{RIEM}'s configuration spaces.
$\Pi_{\sfZ}$ are then the momenta conjugate to the matter variables $\uppsi^{\sfZ}$. 

\m  

\n{\bf Example 1)} The Einstein--scalar case is useful for Cosmology, including this Series of Articles's main Minisuperspace model and perturbations thereabout.
Here
\beq
{\cal S}^{\sG\sR\mbox{-}\phi}_{\sR\se\sll}  \es  \iint_{\sbupSigma} \d^3 x  \, \pa \ms \sqrt{\overline{{\cal R} - 2 \, \slLambda - \half |\pa\phi|^2 - \mV(\phi)}} \mma 
\pa\ms  \es  ||\pa_{\underline{\sF}}(\mh, \phi)||_{\biscM(\sbh)} \m , 
\label{S-BSW-Type}
\eeq
for $\bicM$ as given in Appendix \ref{RIEM}.
Moreover,  
\beq
\scH  \:=  {||\bp||_{\sbN}}^2  + \half \pi_{\phi}^2 - \overline{{\cal R} - 2 \, \slLambda - \half |\pa\phi|^2 - \mV(\phi)}  
      \es  0                                                                                                                                                                  \m ,
\eeq
\beq
\bscM_i \:=  - 2 \, {\cal D}_j {\mp^j}_i 
          \es  - \pi_{\phi}  \phi_{,i}                                                                                                                                           \m , 
\eeq
\beq
\lmt^{\se\sm}_{\sR\si}(\uix)  \es  \lE^{\prime}_{\usF \, \in \, Diff(\sbupSigma)}  \frac{  \int ||\pa_{\underline{\sF}}(\mh, \phi)||_{\biscM(\sbh)}  }  
                                                                                   {  \overline{{\cal R} - 2 \, \slLambda - \half |\pa\phi|^2 - \mV(\phi)}  }                 \m ,
\eeq
$$ 
{\cal D}_j \left\{ \sqrt{\frac{{\cal R} - 2 \, \slLambda - \half |\pa\phi|^2 - \mV_{\phi}}
                        {\{\mh^{ac}\mh^{bd} - \mh^{ab}\mh^{cd}\} \{ \pa{\mh}_{ab} - 2 \, {\cal D}_{(a}\pa\mF_{b)} \}
						                                         \{ \pa{\mh}_{cd} - 2 \, {\cal D}_{(c}\pa\mF_{d)} \} + \half |\pa_{\usF} \phi|^2 }} 
\right.
$$
$$
\times \left.
\{\mh^{jk} \delta^{l}_{i} - \delta^{j}_{i}\mh^{kl}\}\{\pa{\mh}_{kl} - 2 \, {\cal D}_{(k}\pa{\mF}_{l)} \}\right\}  \es 
$$
\beq
- \sqrt{\frac{    {\cal R} - 2 \, \slLambda - \half |\pa\phi|^2 - \mV_{\phi}   }
             {    \{ \mh^{ac} \mh^{bd} - \mh^{ab} \mh^{cd} \}\{ \pa{\mh}_{ab} - 2 \, {\cal D}_{(a}\pa\mF_{b)} \}
			                                                 \{ \pa{\mh}_{cd} - 2 \, {\cal D}_{(c}\pa\mF_{d)} \} + \half |\pa_{\usF} \phi|^2 }    } 
\pa_{\usF}\phi \, \phi_{,i}                                                                                                                                                   \m . 
\eeq
\n{\bf Example 2)} Einstein--Maxwell Theory has 
\beq
{\cal S}^{\sG\sR\mbox{-}\sA}_{\sR\se\sll}  \es  \iint_{\sbupSigma} \d^3x \, \pa\ms \sqrt{\overline{{\cal R} - 2 \, \slLambda - \mB^2/2}}  \mma 
\pa\ms  \es  ||\pa_{\usF}\mh, \pa_{\usF, \Psi}\mA)||_{\biscM(\sbh)}                                                                                                           \m , 
\eeq
\beq
\scH  \:=  \mN_{ijkl}\mp^{ij}\mp^{kl} + \half \uppi_i\uppi^i - \overline{  {\cal R} - 2 \, \slLambda -  \half \mB^2 }  
      \es  0                                                                                                                                                                  \m ,
\eeq
\beq
\scG  \:=  \pa_i \uppi^i  
      \es        0                                                                                                                     \m , 
\eeq
\beq
\scM_i  \:=  - 2 \, {\cal D}_{j}{\mp^j}_i  
        \es  - \{\uppi \cr \mB\}_i  - \mA_i \scG                                                                                       \m ,
\eeq
\beq
\lmt^{\se\sm}_{\sR\si}(\uix)  \es  
\mbox{\Large E}^{\prime}_{\usF , \Psi \in \, Diff(\sbupSigma)\times U(1)(\sbupSigma)}       
\int \left. \sqrt{||\pa_{\usF}\bh||_{\sbM}^2 + \half |\pa_{\usF, \Psi}\mA|^2} \right/   \sqrt{     \overline{{\cal R} - 2 \, \slLambda - \half \mB^2}    }  
\m .
\eeq
See \cite{Giu99} for consideration of its Thin Sandwich Problem, which now involves a system of 4 equations for 4 unknowns.

\m

\n As some background on the Einstein--Dirac case, this has an action of the schematic form 
\beq
{\cal S}  \es  \iint_{\sbSigma}\d^3x \big\{\sqrt{2}\sqrt{\cal W}\pa \ms_{\sq\su\sa\sd} + \pa\ms_{\sll\si\sn} \big\}  \m .
\label{S-ED}
\eeq
I.e.\ a locally ordered Field Theoretic version of the `Randers type' action \cite{Van, ARel2} (a subcase of Jacobi--Synge action).
Moreover, the species whose changes enter the quadratic and linear arc elements are disjoint: 
only bosonic changes enter the former, and only fermionic ones confer the latter.

\subsection{Configuration spaces for GR alongside minimally-coupled matter}\label{RIEM}

\n{\bf Structure 1} This case is useful through including fundamental-field second-order minimally-coupled bosonic matter. 
The redundant configuration space metric now splits according to the direct sum \cite{Teitelboim}
\beq
\bM = \bM^{\sg\sr\sa\sv} \oplus \bM^{\sm\scc\sm} \m .
\label{grav-oplus-mcm}
\eeq
\n{\bf Notation 1} We use $\uppsi^{\sfZ}$ to denote fundamental-field second-order minimally-coupled bosonic matter, 
consisting of blockwise disjoint species $\uppsi^{\sfz}$, with $\fz \in \fZ$.

\m 

\n{\bf Remark 1} It is usually additionally assumed that $\bM$ is independent of the matter fields.  
This example covers e.g.\ minimally-coupled scalars, Electromagnetism, Yang--Mills Theory and scalar Gauge Theories, in each case coupled to GR.

\m

\n{\bf Structure 2} In the case of a minimally-coupled scalar field, we denote the corresponding configuration space by 
\be 
\RIEM(\bupSigma) \m . 
\ee.
\n{\bf Notation 2} We generally use the capped version of a GR configuration space to denote further inclusion of a minimally-coupled scalar field.

\m

\n{\bf Structure 3} The (undensitized) metric on this takes the blockwise form 
\be 
\bicM(\bh) := \mbox{\Huge{(}} \stackrel{\mbox{\normalsize  1 \, \, \, \, \, 0}} {\mbox{\normalsize 0  \,\, $\bM(\bh)$}} \mbox{\Huge{)}}
\ee 
for $\bM(\bh)$ the GR configuration space metric itself.
[This immediately extends to the case of $N$ minimally-coupled scalar fields.]

\subsection{Example 4) Strong Gravity}\label{Strong-Gravity}

For now, by Strong Gravity we mean the strong-coupled limit of GR \cite{I76}.
Via the Belinskii--Khalatnikov--Lifshitz Conjecture \cite{BKL}, 
this is widely believed to be applicable to the primordial-cosmology universe near a singularity.
Strong Gravity exists in both geometrodynamical and metrodynamical forms, with two and five degrees of freedom per space point respectively.
The geometrodynamical case follows from
\beq
{\cal S}  \es  \sqrt{2}\iint_{\sbSigma}\d^{3}x     \sqrt{- 2\overline{\slLambda}} \, \pa\ms
\label{S-SG-3}
\eeq
for $\pa s$ the usual GR kinetic arc element (\ref{T-Rel-d}), whereas the metrodynamical case has the `bare' kinetic arc element
\beq
\pa\ms  \:=  ||\pa{\bh}||_{\sbM} \m .
\label{T-Rel-d-Metro}   
\eeq
This accounts for the latter's three extra degrees of freedom, since the absence of $\mF^i$ or $\upbeta^i$ means that no momentum constraint $\u{\bscM}$ appears.
\beq
\scH_{\sss\st\sr\so\sn\sg}  \:=  {||\bp||_{\sbN}}^2 + 2 \, \overline{\slLambda} = 0  \m .  
\label{H-Strong}
\eeq



\begin{thebibliography}{99}

\footnotesize

\bibitem{M}                   E. Mach, {\it Die Mechanik in ihrer Entwickelung, Historisch-kritisch dargestellt} (J.A. Barth, Leipzig 1883).    
%
                              An English translation is {\it The Science of Mechanics: A Critical and Historical Account of its Development} 
							  Open Court, La  Salle, Ill. 1960).  
	
\bibitem{Lanczos}             C. Lanczos, {\it The Variational Principles of Mechanics} (University of Toronto Press, Toronto 1949).  

\bibitem{Fox}                 C. Fox, {\it An Introduction to the Calculus of Variations} (Oxford University Press, London 1950). 

\bibitem{CH}                  R. Courant and D. Hilbert, {\it Methods of Mathematical Physics} Vol 1 (Interscience, 1953; (reprinted by Wiley, Chichester 1989).	


\bibitem{ADM}                 R. Arnowitt, S. Deser and C.W. Misner, ``The Dynamics of General Relativity", 
                              in {\it Gravitation: An Introduction to Current Research} ed. L. Witten (Wiley, New York 1962), arXiv:gr-qc/0405109. 

\bibitem{BSW}                 R.F. Baierlein, D.H. Sharp and J.A. Wheeler, 
                              ``Three-Dimensional Geometry as Carrier of Information about Time", Phys. Rev. {\bf 126} 1864 (1962).
							  
\bibitem{WheelerGRT}          J.A. Wheeler, ``Geometrodynamics and the Issue of the Final State", 
                              in {\it Groups, Relativity and Topology} ed. B.S. DeWitt and C.M. DeWitt (Gordon and Breach, N.Y. 1964). 
							  							  
\bibitem{Dirac}               P.A.M. Dirac, {\it Lectures on Quantum Mechanics} (Yeshiva University, New York 1964). 

\bibitem{Battelle}            J.A. Wheeler, in {\it Battelle Rencontres: 1967 Lectures in Mathematics and Physics} 
                              ed. C. DeWitt and J.A. Wheeler (Benjamin, New York 1968).  

\bibitem{DeWitt67}              B.S. DeWitt, ``Quantum Theory of Gravity. I. The Canonical Theory." Phys. Rev. {\bf 160} 1113 (1967).

\bibitem{BO69}                E.P. Belasco and H.C. Ohanian, ``Initial Conditions in General Relativity: Lapse and Shift Formulation", J. Math. Phys. {\bf 10} 1503 (1969). 


\bibitem{BKL}                 V.A. Belinskii, I.M. Khalatnikov and E.M. Lifshitz, ``Oscillatory Approach to a Singular Point in the Relativistic Cosmology", Adv. Phys. {\bf 19} 525 (1970).

\bibitem{DeWitt70}            B.S. DeWitt, ``Spacetime as a Sheaf of Geodesics in Superspace", 
                              in {\it Relativity} (Proceedings of the Relativity Conference in the Midwest, held at Cincinnati, Ohio June 2-6, 1969), 
							  ed. M. Carmeli, S.I. Fickler and L. Witten (Plenum, New York 1970). 

\bibitem{Fischer70}           A.E. Fischer, ``The Theory of Superspace", 
                              in {\it Relativity} (Proceedings of the Relativity Conference in the Midwest, held at Cincinnati, Ohio June 2-6, 1969), 
							  ed. M. Carmeli, S.I. Fickler and L. Witten (Plenum, New York 1970).

\bibitem{Magic}               C.W. Misner, in {\it Magic Without Magic: John Archibald Wheeler} ed. J. Klauder (Freeman, San Francisco 1972).

\bibitem{MTW}                 C.W. Misner, K. Thorne and J.A Wheeler, {\it Gravitation} (Freedman, San Francisco 1973).

\bibitem{RS}                  M. Reed and B. Simon {\it Methods of Modern Mathematical Physics. II. Fourier Analysis, Self-Adjointness} (Academic Press, New York 1975).  

\bibitem{I76}                 C.J. Isham, ``Some Quantum Field Theory Aspects of the Superspace Quantization of General Relativity", Proc. R. Soc. Lond. {\bf A351} 209 (1976).

\bibitem{Kuchar76Other}       K.V. Kucha\v{r}, ``Kinematics of Tensor Fields in Hyperspace. II.", J. Math. Phys. {\bf 17} 792 (1976); 
%
							  ``Dynamics of Tensor Fields in Hyperspace. III.", 801; 
%
							  ``Geometrodynamics with Tensor Sources IV'', {\bf 18} 1589 (1977).
			
\bibitem{Christodoulou3}      D. Christodoulou, ``The Chronos Principle and the Interactions of Fields of Spin 0 and 1", 
                              in {\it Trieste 1975, Proceedings, Marcel Grossmann Meeting On General Relativity} (Oxford University Press, Oxford 1977). 
			
\bibitem{Arnold}              V.I. Arnol'd, {\it Mathematical Methods of Classical Mechanics} (Springer, New York 1978).  	
							  
					
\bibitem{Teitelboim}          C. Teitelboim, ``The Hamiltonian Structure of Spacetime", 
                              in {\it General Relativity and Gravitation: One Hundred Years after the Birth of Albert Einstein} Vol 1 ed. A. Held (Plenum Press, New York 1980).

\bibitem{K81}                 K.V. Kucha\v{r}, ``Canonical Methods of Quantization", in {\it Quantum Gravity 2: a Second Oxford Symposium} 
                              ed. C.J. Isham, R. Penrose and D.W. Sciama (Clarendon, Oxford 1981).
							  							  
\bibitem{AMP}                 Y. Choquet-Bruhat, C. DeWitt-Morette and M. Dillard-Bleick, {\it Analysis, Manifolds and Physics} Vol. 1 (Elsevier, Amsterdam 1982).  

\bibitem{Hamilton82}          R.S. Hamilton, ``The Inverse Function Theorem of Nash and Moser", Bull. Amer. Math. Soc. {\bf 7} 65 (1982). 

\bibitem{BrMa}                U. Brechtken-Manderscheid {\it Introduction to the Calculus of Variations} (T.J. Press, Padstow, Cornwall 1991: English translation of 1983 German text).  

\bibitem{TVS}                 K. J\"{a}nich, {\it Topology} (Springer--Verlag, New York 1984).

\bibitem{Wald}                R.M. Wald {\it General Relativity} (University of Chicago Press, Chicago 1984).
	
\bibitem{DS88}                N. Dunford and J.T. Schwarz {\it Linear Operators. Part I. General Theory}, (Wiley, Hoboken N.J. 1988).
		
\bibitem{BY1}                 J.D. Brown and J.W. York, ``Jacobi's Action and The Recovery Of Time In General Relativity", Phys. Rev. {\bf D40} 3312 (1989).
	
			
\bibitem{K91}                 K.V. Kucha\v{r}, ``The Problem of Time in Canonical Quantization", in {\it Conceptual Problems of Quantum Gravity} ed. 
                              A. Ashtekar and J. Stachel (Birkh\"{a}user, Boston, 1991).  
			
\bibitem{K92}                 K.V. Kucha\v{r}, ``Time and Interpretations of Quantum Gravity", 
                              in {\it Proceedings of the 4th Canadian Conference on General Relativity and Relativistic Astrophysics} 
                              ed. G. Kunstatter, D. Vincent and J. Williams (World Scientific, Singapore 1992).  
							  							  
\bibitem{I93}                 C.J. Isham, ``Canonical Quantum Gravity and the Problem of Time",
                              in {\it Integrable Systems, Quantum Groups and Quantum Field Theories}  
                              ed. L.A. Ibort and M.A. Rodr\'{\i}guez (Kluwer, Dordrecht 1993), gr-qc/9210011.

\bibitem{BF}                  R. Bartnik and G. Fodor, ``On the Restricted Validity of the Thin-Sandwich Conjecture", Phys. Rev. {\bf D48} 3596 (1993). 

\bibitem{Giu95b}              D. Giulini, ``What is the Geometry of Superspace?", Phys. Rev. {\bf D51} 5630 (1995), gr-qc/9311017. 
 		
\bibitem{B94I}                J.B. Barbour, ``The Timelessness of Quantum Gravity. I. The Evidence from the Classical Theory",               
                              Class. Quant. Grav. {\bf 11} 2853 (1994).

\bibitem{Lang95}              S. Lang, {\it Differential and Riemannian Manifolds} (Springer, New York 1995). 
							  
\bibitem{FodorTh}             G. Fodor, ``The Thin Sandwich Conjecture" (Ph.D. Thesis, Budapest	1995).						  
							  							  
\bibitem{FM96}                A.E. Fischer and V. Moncrief, ``A Method of Reduction of Einstein's Equations of Evolution 
                              and a Natural Symplectic Structure on the Space of Gravitational Degrees of Freedom", Gen. Rel. Grav. {\bf 28}, 207 (1996).

\bibitem{Giu99}               D. Giulini, ``The Generalized Thin-Sandwich Problem and its Local Solvability", J. Math. Phys. {\bf 40} 1470 (1999), gr-qc/9805065.
							  							  
\bibitem{K99}                  K.V. Kucha\v{r}, ``The Problem of Time in Quantum Geometrodynamics",  in {\it The Arguments of Time}, ed. J. Butterfield 
                              (Oxford University Press, Oxford 1999).
							  
\bibitem{Kendall}             D.G. Kendall, D. Barden, T.K. Carne and H. Le, {\it Shape and Shape Theory} (Wiley, Chichester 1999).  

	
\bibitem{RWR}                 J.B. Barbour, B.Z. Foster and N. \'{o} Murchadha, ``Relativity Without Relativity", 
                              Class. Quant. Grav. {\bf 19} 3217 (2002), gr-qc/0012089.

							  						  
\bibitem{AB}                 E. Anderson and J.B. Barbour, ``Interacting Vector Fields in Relativity without Relativity", Class. Quant. Grav. {\bf 19} 3249 (2002), gr-qc/0201092. 
						  
\bibitem{ABFO}                E. Anderson, J.B. Barbour, B.Z. Foster and N. \'{o} Murchadha, ``Scale-Invariant Gravity: Geometrodynamics", 
                              Class. Quant. Grav. {\bf 20} 157 (2003), gr-qc/0211022.
							  						 
\bibitem{Van}                E. Anderson, ``Variations on the Seventh Route to Relativity", Phys. Rev. {\bf D68} 104001 (2003), gr-qc/0302035. 

\bibitem{KieferBook}          C. Kiefer, {\it Quantum Gravity} (Clarendon, Oxford 2004).  

\bibitem{Lan2}                E. Anderson, `` Does Relationalism Alone Control Geometrodynamics with Sources?", in {\it Classical and Quantum Gravity Research}, 
                              ed. M.N. Christiansen and T.K. Rasmussen (Nova, New York 2008), arXiv:0711.0285.  

\bibitem{FEPI}                E. Anderson, ``New Interpretation of Variational Principles for Gauge Theories. I. Cyclic Coordinate Alternative to ADM Split", 
                              Class. Quant. Grav. {\bf 25} 175011 (2008), arXiv:0711.0288.

\bibitem{PPSCT}               E. Anderson, ``Relational Motivation for Conformal Operator Ordering in Quantum Cosmology", Class. Quant. Grav. {\bf 27} 045002 (2010), arXiv:0905.3357.  

\bibitem{Giu09}               D. Giulini, ``The Superspace of Geometrodynamics", Gen. Rel. Grav. {\bf 41} 785 (2009) 785, arXiv:0902.3923.  
	
	
\bibitem{APoT}                E. Anderson, in {\it Classical and Quantum Gravity: Theory, Analysis and Applications} 
                              ed. V.R. Frignanni (Nova, New York 2011), arXiv:1009.2157. 
							  
\bibitem{Lee1}                J.M. Lee, {\it Introduction to Topological Manifolds} (Springer, New York 2011).
							  
\bibitem{FileR}               E. Anderson, ``The Problem of Time and Quantum Cosmology in the Relational Particle Mechanics Arena", arXiv:1111.1472.  

\bibitem{APoT2}               E. Anderson, Invited Review in Annalen der Physik, {\bf 524} 757 (2012), arXiv:1206.2403.   

\bibitem{ARel2}              E. Anderson, ``Machian Time Is To Be Abstracted from What Change?", arXiv:1209.1266.  

\bibitem{AM13}               E. Anderson and F. Mercati, ``Classical Machian Resolution of the Spacetime Construction Problem", arXiv:1311.6541. 
 
\bibitem{AObs}                E. Anderson, ``Beables/Observables in Classical and Quantum Gravity", SIGMA {\bf 10} 092 (2014), arXiv:1312.6073. 
							  							  							  
\bibitem{APoT3}               E. Anderson, ``Problem of Time and Background Independence: the Individual Facets", arXiv:1409.4117.    

\bibitem{TRiPoD}              E. Anderson, ``TRiPoD (Temporal Relationalism implementing Principles of Dynamics)", arXiv:1501.07822.  
	
\bibitem{Giu15}               D. Giulini, ``Dynamical and Hamiltonian formulation of General Relativity", 
                              Chapter 17 of {\it Springer Handbook of Spacetime} ed. A. Ashtekar and V. Petkov (Springer Verlag, Dordrecht, 2014), arXiv:1505.01403.   
		
\bibitem{ABook}               E. Anderson, {\it Problem of Time. Quantum Mechanics versus General Relativity}, (Springer International 2017) Found. Phys. {\bf 190};  
                              free access to its extensive Appendices is at https://link.springer.com/content/pdf/bbm
														  									   					
\bibitem{ALett}  	          E. Anderson, ``A Local Resolution of the Problem of Time", arXiv:1809.01908.     

\bibitem{MBook}               F. Mercati {\it Shape Dynamics: Relativity and Relationalism} (O.U.P., New York 2018). 
							
\bibitem{A-CBI}               E. Anderson, ``Shape Theory. III. Comparative Theory of Backgound Independence", arXiv:1812.08771. 

\bibitem{I}                   E. Anderson,  ``A Local Resolution of the Problem of Time. I. Introduction and Temporal Relationalism", arXiv:1905.06200.  

\bibitem{II}                  E. Anderson,  ``A Local Resolution of the Problem of Time. II. Configurational Relationalism", arXiv:1905.06206.  

\bibitem{III}                 E. Anderson,  ``A Local Resolution of the Problem of Time. III. The other aspects piecemeal", arXiv:1905.06212.  

\bibitem{IV}                  E. Anderson,  ``A Local Resolution of the Problem of Time. IV. Quantum outline and piecemeal Conclusion", arXiv:1905.06294.  

\bibitem{V}                   E. Anderson,  ``A Local Resolution of the Problem of Time. V. Combining Temporal and Configurational Relationalism for Finite Theories", 
                              arXiv 1906.*****.  

\bibitem{VII}                 E. Anderson,  ``A Local Resolution of the Problem of Time. VII. Constraint Closure", arXiv 1906.*****.


\bibitem{IX}                  E. Anderson,  ``A Local Resolution of the Problem of Time. IX. Spacetime Reconstruction", arXiv 1906.*****.


\bibitem{XI}                  E. Anderson,  ``A Local Resolution of the Problem of Time. XI. Slightly Inhomogeneous Cosmology",  forthcoming.

\bibitem{XII}                 E. Anderson,  ``A Local Resolution of the Problem of Time. XII. Foliation Independence",  forthcoming.


\bibitem{XIV}                 E. Anderson,  ``A Local Resolution of the Problem of Time. XIV. Supporting Account of Lie's Mathematics", forthcoming.
								
\bibitem{Forth}               E. Anderson, forthcoming.   
		  							 
\end{thebibliography}
\end{document}